\documentclass[UTF8,a4paper]{article}
\usepackage{amsmath}
\usepackage{bm}
\usepackage{graphicx}
\usepackage{subfigure}
\usepackage{ mathrsfs }
\usepackage{epstopdf}
\usepackage{ marvosym }
\usepackage{ tipa }
\usepackage{geometry}
\usepackage{ wasysym }
\usepackage{ulem}
\usepackage{indentfirst}
\usepackage{amsfonts,amssymb,dsfont}
\usepackage{accents}

\usepackage{setspace}
\usepackage{threeparttable}
\usepackage{amsmath,mathtools,amsthm}
\usepackage{array}
\usepackage{extarrows}
\usepackage{upgreek}

\makeatletter
\renewcommand*\env@matrix[1][\arraystretch]{%
	\edef\arraystretch{#1}%
	\hskip -\arraycolsep
	\let\@ifnextchar\new@ifnextchar
	\array{*\c@MaxMatrixCols c}}
\makeatother

\usepackage[pdfstartview=XYZ,
bookmarks=true,
colorlinks=true,
linkcolor=blue,
urlcolor=blue,
citecolor=blue,
pdftex,
bookmarks=true,
linktocpage=true, 
hyperindex=true
]{hyperref}
\usepackage{orcidlink}

	\title{\bf Statistic physics in bound state in 3D fermi gas system}
	\author{Chen-Huan Wu 
	\orcidlink{0000-0003-1020-5977} 
		\thanks{chenhuanwu1@gmail.com}
		\\College of Physics and Electronic Engineering, Northwest Normal University, Lanzhou 730070, China}
	
\begin{document}
		
	\maketitle

	\begin{abstract}
			The bound state is treated as a long-lived quasiparticle 
			with slow momenta and current relaxation in fermi liquid phase.
In this paper,
			we discuss the realization of ensemble behavior in a 3D fermi gas system in non-fermi liquid phase.
			We reveal the relation between UV cutoff of relative momentum $\Lambda_{q}$ and its ensemble behavior.
			The  behavior of a bound state system has rarely been investigated before,
			and the relation between the scattering momentum and the  physics has not yet been explored before (to best of our knowledge).
			We found that the cutoff $\Lambda_{q}$ directly related to the distribution and statistical variance of coupling term,
			which becomes Gaussian variable (or Chi-square variable) in  limit (i.e., with a large step number of a fractional distance $\Lambda_{q}^{-1}\rightarrow\infty$).
			Also, we show that the different $\Lambda_{q}^{-1}$ lending support to different phases,
			including non-fermi liquid phase and (disordered) fermi liquid phase,
			which correspond to ill-defined and well-defined bound states, respectively.
			The pair condensation induced by local coupling,
			which happen at critical temperature, would suppresses the non-fermi liquid for constant on-site coupling (it is not the case when the coupling be variant).
Further, when the range of condensation is larger,
the homogeneous character will emerge and the thermalization can be observed
where eigenstates with higher energy has lower overlap with the low-entangled states.
			\\

	\end{abstract}
\begin{small}

		\section{Introduction}

		In conformal limit 
with a U(1) global symmetry in nonperturbed  mode,
 there is a spontaneous symmetry breaking due to the 
		finite expectation $\langle b\rangle$, where $b$ is the boson operator induced by the charge or spin fluctuation,
		the many-body spectrum is gapped out but may accompanied by a gapless Goldstone mode due to the preserved (subsystem) symmetries
		(if any), which is stable against to the phase fluctuation of corresponding order parameter.
		The U(1) gauge symmetry of four-point term with four-fermion interaction will be broken by the hopping term which lacks its hermitian conjugate.
		While for the two-fermion interacting system,
		the $U(1)\times U(1)$ can be preserved as long as the hopping does not contribute to the 
		inter-mode transition (e.g., in the SU(2) symmetries case),
		and thus the four-point term is unstable against the perturbation of bilinear term.
		
		Different to the standard  coupling which is nonlocal and dense,
		the bound state coupling is constant and short-ranged and thus the many-body effect is perturbative.
		While the  mode has an nonperturbative many-body effect in Wigner-Dyson random matrix esemble,
		which can not be fully captured by two-point correlation (single-particle function),
		and thus the functional-derivative approach\cite{Allen S} wound not be useful here,
		due to the randomness of  coupling and disorders.
		However, we will see that,
		after considering the normally distributed interactions among fermion indices and momenta,
		the coupling which is considered to be short-ranged (and even approximately the contact type) 
		can still be able to creates the spectrum
		similar to the one with short-range spectral correlation of  model \cite{Sedrakyan T A},
		and also, the coupling decays exponentially with real space distance 
		(or approximately decays exponentially with bandwidth or coherence scale in momentum space),
		which is similar to the exponentially decay of  mode self-energy in long-time limit
		(low-energy boson excitation) with charging process.
		That reveals the possibility to realize in  physics in bound state system.
	For non-fermi-liquid system,
the ensemble behavior may emerges in bound state system in the range $U/N \ll\omega_{c}\ll \omega\ll g_{q}\ll N$ of frequency space,
		where $\omega_{c}=W^{2}/g_{q}$ is coherence scala with spectrum bandwidth $W$.
		Such a limit can be obtained in real space by replacing the frequency with lattice spacing as given in Ref.\cite{Lantagne-Hurtubise,Pikulin D I}.

Incoherent non-fermi liquid in the gapless critical metal phase
		where the stable mode (boson excitation insteads of quasiparticle) 
		can remains gapless even without turnning to the quantum critical point (and thus without the condensation)
		due to the preserved symmetries,
		and thus contributes to the many-body chaotic spectrum.
		Since the bound state is a kind of quasiparticle or excitation which can exists in both the fermi liquid\cite{Alexandrov A S}
		and non-fermi liquid states.
		In fermi liquid system, the bound state can be detected as a coherence peak emerges in a broad incoherent background of fermi sea,
		and the zero momentum bound state becomes stable and long-lived quasiparticle at zero-temperature limit.
		As the temperature increases, it may be decayed into particle-hole excitations.
		Even a Cooper-like bound stateic pair can exist in bound stateic fermi liquid of many-body system 
		when it satisfies $g\le W<\omega$ ($g$ is the bound stateic coupling, $W$ is the bound state bandwidth, $\omega$ is the fermionic frequency).
		That is to say, the superconducting ordering is possible to exists in bound stateic fermi liquid in the 
		weak-interacting and high frequency (UV) limit.
		For bound state as a quasiparticle in fermi liquid state,
		it will has well-define momenta and can be described by the purely local description in momentum space.
		That makes it roubust against the short-range perturbations, like the Coulomb repulsive interaction and the 
		short-range hopping.
		While the  mode emerges in the range $g\gg \omega\gg W^{2}/g$,
		which requires strong interaction, and thus can be extended to zero temperature as the bandwidth is finite.
		The emergence of incoherent spectral function in a local form signals the appearance of  mode.
		Similar to the bound state, the  model can be described purely in the orbital space, instead of real space,
		as it mostly being studied in zero-dimensional system (although it can be extended to higher spacial dimension\cite{Gu Y2})
		thus the strong  interaction could be long-range and nonlocal,
		which can efficiently breaks the long-range entanglement of fermi-liquid.
		So what will happen when the coupling becomes Gaussian distributed in a mang-body system,
		and being turned to the range that satisfy the  requirement?
		In this paper, we adopt a orbital space$+$ one-dimensional momentum space description to describe this system,
		where we try to understand the connection and
		competition effect between bound stateic physics and the  physics.
		We found that the emergence of ensemble behavior require small enough momentum $q$, which corresponds to
		to $\Lambda_{q}^{-1}$ much larger than the fermion number $N$.
		Due to the small one-dimensional relative momentum $q$, the  model can be approximately as 1+1-dimensonal model
		(we do not consider the many-body localizaion case in the bulk here).
		By the small $q$ is essential to leads to strong bound stateic coupling
		and the infinite steps for a fractional distance from $q=0$ to $q=\Lambda_{q}$,
		which is totally $\Lambda_{q}^{-1}$.
		This is different to the simple $0+1$ space-time dimensional  models where the disorder average is over the
		consecutive integer values of fermion indices,
		Besides, the small momentum leads to the delocalization in real space,
		i.e., the nonlocal  interaction can further suppresses the fermi liquid.
		This is similar to the effect of long-range (weakly screened when close to half-filling)
		Coulomb interaction, which can turn the fermi liquid to non-fermi liquid.
		In the mean time, the robustness of bound state in fermi-liquid to the short range hopping as well as the bilinear chemical potential term
		may help to stablizing the emergent  mode, before the long-range entangment of fermi-liquid is completely destroyed by  coupling.
		
		In the presence of pair condensation, it usually competes with the  non-fermi liquid phase,
		with the coherence appears with finite pairing order parameter
		and for temperature lower than the critical one.

		Note that for many-body localized state,
		the ratios of adjacent level spacings follows the Poisson level statistic,
		while that in random local field state
		follows the Wigner-Dyson level statistic according to eigenstate thermalization hypothesis\cite{You Y Z}.
	
The emerging bound state is directly evidenced in the gapless region,
which corresponds to the negative interaction anisotropy in interaction anisotropy
in XXZ Heisenberg chain.
	
In the presence of random local field with defined Gaussian variables $\Phi_{i}$ and $\Phi^{q}$
		(wave functions about the fermion index and relative momentum, respectively),
		the thermalization and localization can happen in the mean time to a single fermion,
		which is different to the case described in Ref.\cite{You Y Z}.
		And in this case, the fermions can be localized by the on-site interaction and thermalized due to the vanishing level spacing,
		by virtue of random on-site potential.
		
		Once the anomalous components emerge, i.e., the pairing order parameter is nonzero,
		the level statistic should change from Gaussian unitary ensemble (GUE) to Gaussian orthogonal ensemble (GOE),
		where GOE has a level repulsion larger than that of GUE in the small level spacing limit during the level statistic.
		Both the GUE and GOE follow the Wigner-Dyson distribution.
		However, when the pair condensation happen, in which case the system changes
		from the volume law phase (with eigenstate thermalization hypothesis) where the 
		entanglement entropy scales linearly with the bipartition size,
		to the area law phase where the 
		entanglement entropy scales independent of the bipartition size and with short range entanglement\cite{Levy R}.
		This phase transition can be realized by a many-localized field with strong disorder
		which can induces random but short-ranged and sparse interactions\cite{You Y Z},,
		or a random quantum circuit with local measurement\cite{Skinner B}.
		Note that here the pair condensation has similar effect with the product of two bilinear term in the unthermalized  system\cite{Bi Z2}.
		In the extreme case where the eigenvalue splitting is being maximized,
		the off-diagonal long-range order emerges.
		
		Our main conclusion is that, the  physics can be realized in the small bound stateic mmentum $\Lambda_{q}$ limit,
		which is much smaller than the randomly distributed interactions,
		and in the mean time, the bound state can exists.
		But once $\Lambda_{q}=0$,
		the system reduces to the product of two $SYK_{q=2}$ modes
		due to the flat non-fermi-liquid spectral function.
		Note that, although the tradictional  physics require the system to be momentum-independent,
		i.e., in zero-dimensional space,
		which can be experimentally realized under strong magnetic field\cite{Chen A} or in a Kagome-type optical lattice\cite{Wei C}
		to obtain the flat levels,
		we here provide a route to realize the  physics without completely removing the momentum-dependence,
		but transform the momentum-dependence to the fermion index-dependence with a certain constrain.
		We will discuss in detail about this constrain and to what extent this model can realize the  physics in this paper.
Our study also reveals the strong suppression effect of strong pairing coupling to the quantum chaos,
and it can be realized experimentally in terms of the single-qubit interferometry\cite{Cao}.
		
		\section{Effective mass}
		
				Firstly we discuss the electronic properties of bound state and the role played by the relative momentum $q$
		during the scattering.
		When the formation of bound state\cite{Kohstall C} does not contains the component which has a macroscopic amount,
		like the phonons or photons (in a cavity),
		the total momentum of the weakly-coupled propagating pair is conservative,
		and thus the transport relaxation time of bound state reads
		$1/\tau\sim \Sigma^{(1)}=ng_{b}$ where $\Sigma^{(1)}$ is the
		first order contribution to self-energy.
		$g_{b}$ is the bare coupling strength which equals
		to the vacuum scattering matrix.
		For temperature higher than the critical one ($T_{c}=W^{2}/g_{b}$ with $W$ the bandwidth)
		the non-fermi-liquid feature,
		which is hidden by the instability at low temperature,
		will leads to incoherence between the electronic excitations.
		Note that the finite-temperature also affects the resonance structure of the bound state spectrum\cite{Field B}.
		Besides, the strong coupling in strongly correlated metals will
		largely reduce the critical temperature, and thus lead to the
		coherence-to-incoherence crossover,
		which also happen in the Dirac/Weyl semimetallic states.
		
		By writting the isotropic dispersions of impurity (with spin $\sigma$) and majority (with spin $\sigma'$) particles 
		as (before collision) $\varepsilon_{p\sigma}=p^{\alpha}$ and $\varepsilon_{k\sigma'}=k^{\beta}$, respectively,
		the Green's function in fermi-liquid phase are 
		$G_{p\sigma}^{-1}=iZ^{-1}\omega-\varepsilon_{p\sigma}$
		and 
		$G_{q\sigma'}^{-1}=iZ^{-1}\Omega-\varepsilon_{q\sigma'}$, respectively,
		where $Z=1-(g_{q}D_{0})^2$ is the quasiparticle residue in weak-coupling limit\cite{Chowdhury D},
		and we define $g_{q}$ as the coupling and $D_{0}$ as the bare density-of-states.
		The Chevy-type variational ansatz for a mobile impurity with momentum $p$ dressed by one
		electron-hole pairs (excitations)
		is
		\begin{equation} 
			\begin{aligned}
				|\psi\rangle=\psi_{0}b^{\dag}_{p\sigma}|0\rangle_{\sigma'}
				+\sum_{q}\psi_{kq}c_{p+q-k,\sigma}^{\dag}
				c_{k,\sigma'}^{\dag}|0\rangle_{\sigma'},
			\end{aligned}
		\end{equation}
		where $|0\rangle_{\sigma'}=\Pi_{k<k_{F}}c_{k\sigma'}^{\dag}|{\rm vac}\rangle$ is the ground state of majority particles.
		$c^{\dag}_{k\sigma'}$ is the creation operator of the 
		excited particle with momentum $k$,
		and $c_{q\sigma'}$ is the the annihilation operator of the hole at momentum $q$.
		Here $k>k_{F}$ is to make sure the particles are excited out of the fermi surface.

		Due to the interaction, the impurity (fermion) will dynamically obtain a finite fermion mass,
		which is possible to leads to the charge-density wave (CDW) state.
		The excitonic mass generation is a nonperturbative result,
		and the dynamically generated mass term reads
		\begin{equation} 
			\begin{aligned}
				m(p,\omega)=\int_{q}\int\frac{d\Omega}{2\pi}\frac{m_{0}}{\Omega^{2}+q^{2}+m_{0}^{2}}T(p-q,\omega-\Omega),
			\end{aligned}
		\end{equation}
		where $m_{0}$ is the bare fermion mass in the absence of bound stateic coupling.
		For large order $\beta$, the low-energy DOS
		can be treated as a constant $m$. 
		Firstly, we discuss the $\Pi=0$ case,
		\begin{equation} 
			\begin{aligned}
				m(p,\omega)
				=&\int_{q}\int\frac{d\Omega}{2\pi}\frac{m_{0}}{\Omega^{2}+q^{2}+m_{0}^{2}}g_{q}\\
				=&\frac{1}{2}\int_{q}\frac{m_{0}}{\sqrt{q^{2}+m_{0}^{2}}}g_{q}\\
				=&\frac{1}{4\pi} g_{q} {\rm ln}[1 + \frac{2 \Lambda_{q} (\Lambda_{q} + \sqrt{m_{0}^2 + \Lambda_{q}^2})}{m_{0}^2}] m_{0}.
			\end{aligned}
		\end{equation}
		
		In long wavelength limit, the contribution from electron-hole excitation vanishes, and
		we have
		\begin{equation} 
			\begin{aligned}
				m(\omega)
				=&\int\frac{d\Omega}{2\pi}\frac{m_{0}}{\Omega^{2}+m_{0}^{2}}T_{ee}(\omega-\Omega)\\
				=&\int\frac{d\Omega}{2\pi}\frac{m_{0}}{\Omega^{2}+m_{0}^{2}}[g_{q}^{-1}-D_{0}(1-\frac{\omega-\Omega-iZ}{\omega-\Omega})]^{-1}\\
				=&-\frac{(g_{q} \pi (m_{0} + i\omega)}{g_{q} Z + m_{0} +i \omega)},
			\end{aligned}
		\end{equation}
		where $m_{0},g_{q}<0$.
		In fermi-liquid phase with $\omega\rightarrow 0$ limit,
		it is possible to enters into the hydrodynamic regime when the bound stateic interaction scattering rate is large (strong interacting fermi liquid)
		$\tau\propto g_{q}^{-1}$.
		And in this case the coupling $g_{q}$ is perturbative.
		Since here the Lorentz invariance is broken,
		the dependence of mass term on momentum and frequency (of impurity) can be treated 
		separately.
		We found from this expression that the mass term
		is proportional to the impurity self-energy and inversely proportional to the
		eigenenergy of majority particle.
		This is similar to mass enhancement due to the electron-phonon coupling
		as described by the Eliashberg theory
		Since we apply the gapless dispersion for both the impurity and majority particles in the begining,
		the mass enhancement here (due to the interaction effect) is also the final dispersion gap.
		In high-frequency (UV) limit,
		the mass tends to zero,
		which implies that the irrelavance of interaction effect and the bound state dynamic.

		While in the instantaneous approximation,
		the momentum dependence of mass can be obtained as
		\begin{equation} 
			\begin{aligned}
				m(p)
				=&\int\frac{d^{d}q}{(2\pi)^{d}}\frac{m_{0}}{q^{2}+m_{0}^{2}}T(p-q)\\
				=&\int\frac{d^{d}q}{(2\pi)^{d}}\frac{m_{0}}{q^{2}+m_{0}^{2}}  [g_{q}^{-1}-D_{0}(1-\frac{-iZ}{Zv_{F}(p-q)})]^{-1}
				+\int\frac{d^{d}q}{(2\pi)^{d}}\frac{m_{0}}{q^{2}+m_{0}^{2}}  [g_{q}^{-1}-ZD_{0}]^{-1}.
			\end{aligned}
		\end{equation}
		This can be solved analytically in one-dimension space as
		\begin{equation} 
			\begin{aligned}
				m(p)
				=&\frac{g_{q} \pi v_{F} (m_{0} + i p)}{-D_{0} g_{q} + (-1 + D_{0} g_{q}) v_{F} (m_{0} + i p)}
				+\frac{-g_{q} \pi }{1 - g_{q} ZD_{0}},
			\end{aligned}
		\end{equation}
		where $m_{0},g_{q}<0$.
		
		As the dispersion order is treated experimental turnable here,
		we can then obtain a multiple-band system with different bandwidths
		which exhibit marginal fermi liquid feature in a finite temperature range,
		where the transport relaxation time has
		$1/\tau\sim {\rm Im}\Sigma\propto{\rm max}[\varepsilon,T]$
		and the quasiparticle weight has $Z^{-1}\sim{\rm In}\frac{\Lambda}{\varepsilon}$.
		The incoherent part of spectrum becomes dominate over the coherent part 
		($\delta$-function) in this case.
		Note that for impurity scattering,
		when the random quenches of the impurities are taken into account,
		the momentum is no more conservative during the scattering event (due to the finite momentum relaxation),
		and thus the system exhibit non-fermi liquid behavior even at zero temperature ($T=0$)\cite{Buterakos D}.
		Although the impurity we discuss in this paper is mobile,
		the scattering can still be treated as elastic as long as the transferred energy is low enough,
		e.g., much less than the temperature scale.
		When the elastic scattering by quenched impurities is weak enough (low-energy event) and can be treated as a
		perturbation,
		the system can still behaves like a fermi-liquid one,
		e.g., obeys the Wiedemann-Franz law or broadened Drude peak in Ferimi-liquid metal\cite{Mahajan R}.


		\section{Gaussian random coupling in fermi gas}
		
		For fermion operator with scaling dimension $1/4$,
		the four-point term reads
		\begin{equation} 
			\begin{aligned}
				H=\sum_{ijkl}\frac{J_{ijkl}}{(2N)^{3/2}}c_{i}^{\dag}c^{\dag}_{j}c_{k}c_{l},
			\end{aligned}
		\end{equation}
		where $i,j,k,l=1\cdot\cdot\cdot N$. $J_{ijkl}$ is an antisymmetry tensor which follows a Gaussian distribution with zero mean.
		The  coupling satisfies the relation $J_{ijkl}=J_{jilk}=-J_{jikl}=-J_{ijlk}=J_{klij}^{*}=J_{lkij}^{*}$.
		It satisfies the relation after disorder average in Gaussian unitary ensemble 
		\begin{equation} 
			\begin{aligned}
				\label{GUE}
				\frac{J_{ijkl}}{(2N)^{3/2}}\mathcal{O} \rightarrow -\frac{J^{2}}{8N^{3}}2\mathcal{O}^{\dag}\mathcal{O}=-\frac{J^{2}}{4N^{3}}\mathcal{O}^{\dag}\mathcal{O},
			\end{aligned}
		\end{equation}
		where $\mathcal{O}$ is a physical observable quantity
		which consists of fermion operators with indices different to each other.
		similarly, if  we restrict the four indices to the region $i<j; k<l$ (with the corresponding coupling $J_{ij;kl}$),
		we have
		\begin{equation} 
			\begin{aligned}
				\frac{J_{ij;kl}}{(2N)^{3/2}}\mathcal{O} \rightarrow -\frac{J^{2}}{16\times 8N^{3}}2\mathcal{O}^{\dag}\mathcal{O}=-\frac{J^{2}}{(4N)^{3}}2\mathcal{O}^{\dag}\mathcal{O},
			\end{aligned}
		\end{equation}
		since $J_{ij;kl}=J_{ijkl}/4$ (note that this is indeed an approximated result which is accurate only in $N\rightarrow \infty$ limit).
		\footnote{	For fully antisymmetry tensor $J_{ijkl}$ with indices $i,j,k,l=1\cdot\cdot\cdot N$,
			if we do not restrict the indices $i,j,k,l$ are different to each other
			(although that would leads to shift of particle-hole symmetry point away from $\mu=0$),
			then the ratio of 
			$
			\alpha_{q=4}=\lim_{N\rightarrow \infty}\frac{\sum_{ijkl}J_{ijkl}}{\sum_{i<j<k<l}J_{ijkl}}
			$
			can be obtained as
			$
			\alpha_{q=4}^{-1}=\lim_{N\rightarrow\infty}\frac{1}{N^{q}}\sum^{N-(q-1)}_{i=1}(\sum^{N-i-q+1}_{a=0}(a+1)(N-i-q+2-a))
			= \frac{1}{24}=\frac{1}{4!}$.
			That is why the prefactor in Eq.(17) can also be written as $\frac{J^{2}3!}{4!N^{3}}$,
			which is consistent with the prefactors given by us (Eq.(30) and Eq.(33)).
			And similarly,we can obtain
			$\alpha_{q=2}
			=\lim_{N\rightarrow \infty}\frac{\sum_{ij}J_{ij}}{\sum_{i<j}J_{ij}}
			=(\lim_{N\rightarrow \infty}\frac{1}{N^{q}}\sum^{N-(q-1)}_{i=1}(N-i))^{-1}
			=2,\
			\alpha_{q=4}
			=\lim_{N\rightarrow \infty}\frac{\sum_{ijkl}J_{ijkl}}{\sum_{i<j;k<l}J_{ij;kl}}=4$.
			}
		The time-reversal symmetry as well as the particle-hole symmetry (even at zero chemical potential)
		is broken in Gaussian unitary ensemble with finite $N$, but preserved in Gaussian orthogonal ensemble,
		in which case Eq.(\ref{GUE}) becomes 
		\begin{equation} 
			\begin{aligned}
				\frac{J_{ijkl}}{(2N)^{3/2}}\mathcal{O} \rightarrow -\frac{J^{2}}{32N^{3}}\mathcal{O}^{\dag}\mathcal{O}.
			\end{aligned}
		\end{equation}
		In this case the $J_{ijkl}$ is totally antisymmetry and thus the above Hamiltonian is hermitian.

		To mapping to the  basis,
		we introduce the following $s$-wave operators which can expressed in terms of the eigenfunctions of  Hamiltonian
		\begin{equation} 
			\begin{aligned}
				\Delta_{-q\sigma}&=\sum_{p}c_{p-q,\sigma}^{\dag}c_{p\sigma}=\sum_{il}\sum_{p}\phi^{*}_{i}(p-q)\phi_{l}(p)c^{\dag}_{i\sigma}c_{l\sigma},\\
				\Delta_{q\sigma'}&=\sum_{k}c_{p+q,\sigma'}^{\dag}c_{k\sigma'}=\sum_{il}\sum_{k}\phi^{*}_{j}(k+q)\phi_{k}(k)c^{\dag}_{j\sigma'}c_{k\sigma'},\\
			\end{aligned}
		\end{equation}
		where the antisymmetry tensors read
		\begin{equation} 
			\begin{aligned}
				\sum_{p}\phi^{*}_{i}(p-q)\phi_{l}(p)&=\Phi_{il}^{-q},\\
				\sum_{k}\phi^{*}_{j}(k+q)\phi_{k}(k)&=\Phi_{jk}^{q},
			\end{aligned}
		\end{equation}
		and the coupling satisfies
		\begin{equation} 
			\begin{aligned}
				\delta_{il}\delta_{jk}J_{ijkl}=-\sum_{q}^{\Lambda_{q}}\Phi_{il}^{-q}g_{q}\Phi_{jk}^{q}.
			\end{aligned}
		\end{equation}

		Note that the wave functions $\phi_{i},\phi_{j},\phi_{k},\phi_{l}$ are not Gaussian variables,
		but the operators $\Phi_{il},\Phi_{jk}$ are uncorrelated random Gaussian variables with zero mean 
		($\overline{\Phi_{il}}=\overline{\Phi_{jk}}=0$)
		as long as $\Lambda_{q}\neq 0$,
		in which case the index $i$($j$) is completely independent with $l$($k$).
		Otherwise, in the completely  limit $\Lambda_{q}^{-1}\rightarrow \infty$ (and in the mean time, $\Lambda_{q}\rightarrow 0$, which signals the vanishing spacial effect),
		$\Phi_{il}$ and $\Phi_{jk}$ are no Gaussian variables,
		while $(\Phi_{il}\Phi_{jk})$ and $\Phi_{ik}$ ($\Phi_{jl}$) become Gaussian variables.
		Here the coupling $J_{ijkl}$ depends fully on the indices $i,j,k,l$ unlike the following notion $J$,
		which is obtained after the disorder average.
		Then the bound state Hamiltonian can be rewritten as
		\begin{equation} 
			\begin{aligned}
				H_{p}=\frac{g_{p}}{N}\sum_{ijkl}^{N}\Phi_{il}\Phi_{jk}c^{\dag}_{i\sigma}c^{\dag}_{j\sigma'}c_{k\sigma'}c_{l\sigma},
			\end{aligned}
		\end{equation}
		In this case, the bound state coupling is still constant,
		but the wave functions $\Phi_{il}$ and $\Phi_{jk}$ are random independent Gaussian variables.
		This Hamiltonian is similar to the complex  model with a conserved global U(1) charge,
		but with spins.
		
		As we stated in above, the calculations related to the bound state dynamics usually requries momentum cutoff $\Lambda_{q}$.
		The bound state coupling reads (with $E_{b}$ the binding energy and $W$ the bandwidth)
		\begin{equation} 
			\begin{aligned}
				g_{q}^{-1}=-\sum_{kp}^{\Lambda}\frac{1}{E_{b}+\varepsilon_{p}+\varepsilon_{k}+W},
			\end{aligned}
		\end{equation}
		which vanishes in $\Lambda\rightarrow \infty$ limit.
		Similarly, in two space dimension, the bound state corresponds to the pole $q=ia^{-1}$ where $a$ is the scattering length (or scattering amplitude),
		which proportional to the coupling strength,
		and the strongest coupling realized at $q\sim a^{-1}$ while the weakest coupling realized at $qa\ll 1$.
		This is a special property of bound state formation and is important during the following analysis.
		Now we know that $g_{q}$ is inversely proportional to the value of bound stateic exchanging momentum $q$, then
		by further remove the $q$-dependence of bound stateic coupling
		\begin{equation} 
			\begin{aligned}
				\label{couplingintegral}
				g^{-1}=-\sum_{q}^{\Lambda_{q}}\sum_{kp}^{\Lambda}\frac{1}{E_{b}+\varepsilon_{p-q}+\varepsilon_{k+q}+W},
			\end{aligned}
		\end{equation}
		the integral in Eq.(\ref{couplingintegral}) is vanishingly small when $\Lambda_{q}\rightarrow \infty$.
		Note that in the following we may still use $g_{q}$ to denote the coupling to distinct it from the  one.
		In opposite limit,
		when $\Lambda_{q}\rightarrow 0$, both the couplings $J_{ijkl}$ and $g_{q}$ become very strong (thus enters the  regime).
		Similar to the disorder effect from temperature (which is lower than the coherence scale but higher that other
		low energy cutoff) to fermi liquid,
		the fermion frequency can be treated as a disorder to non-fermi liquid  physics,
		(the pure  regime can be realized in $\omega, \omega_{c}\rightarrow 0$ limit and extended to zero temperature)
		thus we can write the essential range of parameter to realizes  physics,
		\begin{equation} 
			\begin{aligned}
				\Lambda_{q}^{-1}\sim \frac{g_{q}}{\omega}v_{F}\gg N,
			\end{aligned}
		\end{equation}
		this is one of the most important result of this paper which relates the bound state physics to the  physics, 
		and in the mean time, it is surely important to keep
		$N\gg U\gg \omega\gg\omega_{c}\gg U/N$.
		Here the $\omega\sim v_{F}q$ plays the role of disorder in frequency space.
		Note that here vanishingly small cutoff in momentum space $\Lambda_{q}$ corresponds to vanishing spacial disorder $\Lambda_{r}$ which is the
		lattice spacing in two-dimensional lattice in real space\cite{Lantagne-Hurtubise}.
		That is, in the presence of short range interaction,
		by reducing the distance between two lattice sites (and thus enlarging the size of hole),
		the size number as well as the coupling is increased.
		Thus in this case the bound state term becomes asymptotically Gaussian distributed due to the virtue of the central limit theorem.
		
		In the $\Lambda_{q}=0$ limit,
		the Hamiltonian can be rewritten in the exact form of $H_{\flat=2}$  mode.
		\begin{equation} 
			\begin{aligned}
				H_{p}=&g_{q}\sum_{ik}\sum_{\sigma,\sigma'}\Phi_{i}c^{\dag}_{i\sigma}c_{i\sigma}\Phi_{k}c^{\dag}_{k\sigma'}c_{k\sigma'}\\
				\equiv &g_{q}\sum_{ik}\sum_{\sigma,\sigma'}b_{i\sigma}b_{k\sigma'},
			\end{aligned}
		\end{equation}
		where $b_{i}=\Phi_{i}c^{\dag}_{i\sigma}c_{i\sigma}$ and $[b_{i},H_{p}]=0$.
		Similar to Eq.(\ref{GUE}),
		after disorder average in Gaussian unitary ensemble (GUE),  for Gaussian variable $\Phi_{il}\Phi_{jk}$,
		we have
		\begin{equation} 
			\begin{aligned}
				\overline{\Phi_{il}\Phi_{jk}}=0,
			\end{aligned}
		\end{equation}
		and the replication process reads
		\begin{equation} 
			\begin{aligned}
				g_{q}\Phi_{il\sigma}\Phi_{jk\sigma'}\mathcal{O}_{il,jk}
				\rightarrow
				-\frac{J^{2}\delta_{ij}\delta_{jk}2\mathcal{O}_{il,jk}^{\dag}\mathcal{O}_{il,jk}}{16N^{2}}
				=-\frac{J^{2}\delta_{ij}\delta_{jk}\mathcal{O}_{il,jk}^{\dag}\mathcal{O}_{il,jk}}{8N^{2}}.
			\end{aligned}
		\end{equation}
		Note that before replication,
		the number of observable $\mathcal{O}$ should equals to the number of Gaussian variables, which is one in the above formula.
		Here the prefactors follows the scheme of $H_{\flat=2}$ ,
		i.e., the above bound state Hamiltonian can be viewed as a $H_{\flat=2}$  model,
		and each fermion operator (with independent indices, like $i$ and $j$ or $l$ and $k$) contributes factor
		\begin{equation} 
			\begin{aligned}
				(2^{\frac{\flat/2+1}{2\flat}}N^{\frac{\flat-1}{2\flat}})^{-1}.
			\end{aligned}
		\end{equation}
		That is to say, since $i=l$ and $j=k$ in $\Lambda_{q}=0$ limit,
		among $c_{i}^{\dag}$ and $c_{l}$ (or  $c_{j}^{\dag}$ and $c_{k}$),
		only one of them contributes to the prefator.
		This is unlike the $H_{\flat=2}$  $\times$ $H_{\flat=2}$  scheme in the finite $\Lambda_{q}$ case as we will introduce in below.
		
		Next we consider the eigenkets of a real symmetric random GOE matrix provided by the basis labelled by scattering momenta $\{|q\rangle \}$.
		Since the eigenstates of random matrices provide orthogonal basis,
		the resulting product $\langle -q|il\rangle$ is Gaussian distributed with zero mean and variance depends on, to leading order, the dimension of the corresponding Hilbert space $D$
		(note that $D=\frac{\Lambda_{q}^{-1}}{2}$), $\overline{\Phi^{-q}_{il}\Phi^{q'}_{jk}}
		=\frac{1}{D}\delta_{q,q'}\delta_{il,jk}$,
		where for each wave function,
		e.g., $\Phi^{-q}_{il}$,
		the corresponding creation and anihilation fermion operators
		carrier the informations as $c^{\dag}_{i}(-q)c_{l}(-q+n\Lambda_{q}^{-1})$
		where $n\Lambda^{-1}_{q}$ denotes arbitary multiples of the flavor number of quantized scattering momentum,
		and such a single density can be viewed as a single-particle kinetic term in terms of the linear response to the perturbation.
In the thermodynamic limit, which corresponds to large size of the system that the normalized eigenstates $| q \rangle$ belong to,
such that $\Lambda_{q}^{-1}\gg N$,
the Gaussian distribution of $\Phi^{-q}_{il}$ results in the emergent finite linear dependence between the vectors
$(\Phi^{-q}_{il}\Phi^{q}_{il},
\Phi^{-q'}_{il}\Phi^{q'}_{il},
\cdots
)^{T}$
and
$(\Phi^{-q}_{jk}\Phi^{q}_{jk},
\Phi^{-q'}_{jk}\Phi^{q'}_{jk},
\cdots
)^{T}$.
This can be related to the central limit theorem,
as well as the thermalization of the basis $\{|q\rangle \}$
with respect to the observable $|il\rangle\langle jk |\delta_{il,jk}$.
		
		While in the finite but small $\Lambda_{q}$ case 
		\begin{equation} 
			\begin{aligned}
				H_{p}=&g_{q}\sum_{il;jk}\sum_{\sigma,\sigma'}
				\Phi_{il\sigma}^{-q}\Phi_{jk\sigma'}^{q}
				c^{\dag}_{i\sigma}c_{l\sigma}c^{\dag}_{j\sigma'}c_{k\sigma'},
			\end{aligned}
		\end{equation}
		where we have
		\begin{equation} 
			\begin{aligned}
				g_{q}\Phi_{il\sigma}^{-q}\Phi_{jk\sigma'}^{q}\mathcal{O}_{ik}^{\dag}\mathcal{O}_{lj}  2\mathcal{O}_{q}^{\dag}\mathcal{O}_{q}
				\rightarrow&
				\frac{J_{il;jk}^{2}\delta_{il}\delta_{jk}\Lambda_{q}^2  4\mathcal{O}_{ik}\mathcal{O}_{ik}^{\dag}\mathcal{O}_{lj}^{\dag}\mathcal{O}_{lj}
					(2\mathcal{O}_{q}^{\dag}\mathcal{O}_{q})^{2}}{16N^{2}}\\
				=&\frac{J_{il;jk}^{2}\delta_{il}\delta_{jk}\Lambda_{q}^2\mathcal{O}_{ik}\mathcal{O}_{ik}^{\dag}\mathcal{O}_{lj}^{\dag}\mathcal{O}_{lj}
					(\mathcal{O}_{q}^{\dag}\mathcal{O}_{q})^{2}}{N^{2}}.
			\end{aligned}
		\end{equation}
		Note that each operator $\mathcal{O}$ must contains $q$ completely independent (uncorrelated) indices,
		and beforce replication, the indices of
		each operator $\mathcal{O}$ must not be completely the same.
		For example, in the finite (although small) $\Lambda_{q}$ case,
		$i$ ($l$) is completely independent of $k$ ($j$), but index $ik$ is not completely uncorrelated with $lj$
		because mapping to their momentum space they have $i-l=k-j$ due to the fixed relative momentum $q$,
		in other word, the mechanism that transforms $i$ to $l$ is the same with that to transform $k$ to $j$,
		thus $ik$ can continuously mapped to $lj$.
		So this is neither the simple $H_{\flat=2}$  scheme or the $H_{\flat=4}$  scheme (which allows the existence of $\mathcal{O}_{ijkl}$),
		but a ($H_{\flat=2}$  $\times$ $H_{\flat=2}$ ) scheme,
		which can be treated as a Chi-square random variables (the product of two independent $H_{\flat=2}$  Gaussian variables).
		After the summation over $q$ is done,
		it becomes ($H_{\flat=2}\times H_{\flat=2} \times H_{\flat'=1}$) scheme.
		We note that in the above expression, after the replication,
		the created new operators $\mathcal{O}$ are all with different imaginary time indices although we omit them here.
		This discussion is also applicable to the  model with SU(M) spin (see Eq.(9) of Ref.\cite{Sachdev S},
		which is also a ($H_{\flat=2}$  $\times$ $H_{\flat=2}$) scheme).

For $H_{\flat=2}$  $\times$ $H_{\flat=2}$,
		each fermion operator contributes factor
		\begin{equation} 
			\begin{aligned}
				(2^{\frac{q/2+1}{2q}}N^{\frac{\flat-1}{2q}})^{-1}[(\Lambda_{\flat}^{-1})^{\frac{1}{2q'}}]^{-1}
				=(2^{\frac{q/2+1}{2q}}N^{\frac{\flat-1}{2q}})^{-1}[\Lambda_{\flat}^{\frac{1}{2q'}}],
			\end{aligned}
		\end{equation}
		where $\flat=2$ since we approximate $i=l(j=k)$ due to the vanishingly small $\Lambda_{q}$,
		and $\flat'=1$ for the antisymmetry tensor $\Phi^{\pm q}$,
where $\overline{\Phi^{-q}}=\overline{\Phi^{q}}=0$.
Both the $\overline{\Phi^{-q}}$ and $\overline{\Phi^{q}}$ are the Gaussian variables and
 $\overline{\Phi^{-q}\Phi^{q}}\neq 0$ (and no more be a Gaussian variable).
		The summation over $q$ indeed follows SYK$_{\flat'=1}$ physics.
		This is obtained by defining a fractional distance between $q\gtrsim 0$ to $q=\Lambda_{q}$ 
		(with totally $\Lambda_{q}^{-1}$ steps and each reads $\delta\Lambda_{q}=\frac{\Lambda_{q}}{\Lambda_{q}^{-1}}
		=\Lambda_{q}^{2}$)
		to carry out the Wigner-Dyson statistics in GUE, although $\Lambda_{q}$ must be rather small 
		to guarantees the  behaviors.
		Also, even we do not approximate $i=l(j=k)$, the summation of $ijlk$ do not follows the $H_{\flat=4}$  physics
		since the four random Gaussian variables $\phi_{\alpha}(\alpha=i,j,k,l)$ are not independent of each other
		(once three of them are identified, the last one will be identified).
		
		In the case of finite (although small) $\Lambda_{q}$ with approximately uncorrelated random Gaussian variables $\Phi_{il}$ and $\Phi_{jk}$
		($\overline{\Phi_{il}}=\overline{\Phi_{jk}}=0$, $\overline{g_{q}\Phi_{il}^{2}}=\overline{g_{q}\Phi_{jk}^{2}}=\frac{J}{2N}2\Lambda_{q}=\frac{J}{N}\Lambda_{q}$),
		we can perform the disorder averages over fermion indices and the $q$ in the same time,
		which leads to the following mean value and variance of Chi-square random variable $g_{q}\Phi^{-q}_{il\sigma}\Phi^{q}_{jk\sigma'}$
		\begin{equation} 
			\begin{aligned}
				\label{Chi}
				\overline{g_{q}\Phi^{-q}_{il}\Phi^{q}_{jk}}=&g_{q}\overline{\Phi^{-q}_{il}}\overline{\Phi^{q}_{jk}}=0,\\
				\overline{(g_{q}\Phi^{-q}_{il}\Phi^{q}_{jk})^{2}}
				=&g_{q}^{2}\overline{(\Phi^{-q}_{il})^{2}}\overline{(\Phi^{q}_{jk})^{2}}
				=\frac{J^{2}4\Lambda_{q}^{2}}{4N^{2}}
				=\frac{J^{2}\Lambda_{q}^{2}}{N^{2}}.
			\end{aligned}
		\end{equation}
		The second line is valid because $(\Phi^{-q}_{il}\Phi^{q}_{jk})^{2}=(\Phi^{-q}_{il})^{2}(\Phi^{q}_{jk})^{2}$ when $\Phi^{-q}_{il}$ is independent of $\Phi^{q}_{jk}$,
		and we assume the variance of $\Phi^{-q}\Phi^{q}$ is zero throught out this paper.
		Note that this only valid in the case that the disorder average over $q$ fermion indices are
		done separately,
		i.e., the degree of freedom of $q$ will not affect the correlation between $\Phi_{il\sigma}$ and $\Phi_{jk\sigma'}$,
		and vice versa.
		Besides, $q$ must be integrated in the same dimension of $N$,
		i.e., one dimension (which is the case we focus on in this paper), and thus $\Lambda_{q}$ is in the same scale with $N$.
		If the $q$-integral is be carried out in $d$-space dimension,
		then the above result becomes 
		\begin{equation} 
			\begin{aligned}
				\overline{(g_{q}^{2}\Phi^{-q}_{il}\Phi^{q}_{jk})^{2}}
				=\frac{J^{2}\Lambda_{q}^{2d}}{N^{2}}.
			\end{aligned}
		\end{equation}
		because the sample number of $q$ is related to spacial dimension $d$.
		
		Next we take the spin degree of freedom into account.
		To understand the effects of perturbation broughted by finite small $q$
		(where the coupling is still approximately viewed as a constant),
		we use the SU(2) basis to deal with the degree of freedom of spin (i.e., of impurity and majority particles),
		$\Phi_{\sigma=\pm}=\frac{1}{\sqrt{2}}(\Phi_{1}\pm i\Phi_{2})$,
		Then we have
		\begin{equation} 
			\begin{aligned}
				&g_{q}\sum_{\sigma,\sigma'=\pm}\sum_{il;jk}\sum^{\Lambda_{q}}_{q}
				\Phi^{-q}_{il\sigma}\Phi^{q}_{jk\sigma'}\\
				=&g_{q}\sum_{q}^{\Lambda_{q}}\sum_{il;jk}
				(\frac{1}{2}(\Phi^{-q}_{il1}+i\Phi^{-q}_{il2})(\Phi^{q}_{jk1}-i\Phi^{q}_{jk2})
				-\frac{1}{2}(\Phi^{-q}_{il1}-i\Phi^{-q}_{il2})(\Phi^{q}_{jk1}+i\Phi^{q}_{jk2}))\\
				=&g_{q}\sum_{q}^{\Lambda_{q}}\sum_{il;jk}
				(i\Phi_{il2}^{-q}\Phi_{jk1}^{q}-i\Phi_{il1}^{-q}\Phi_{jk2}^{q}).
			\end{aligned}
		\end{equation}
		And we obtain the variance
		\begin{equation} 
			\begin{aligned}
				g_{q}\Phi_{il1}^{-q}\Phi_{jk1}^{q}2\mathcal{O}_{il}^{\dag}\mathcal{O}_{il}
				&=g_{q}\Phi_{il2}^{-q}\Phi_{jk2}^{q}2\mathcal{O}_{il}^{\dag}\mathcal{O}_{il}
				=\frac{J}{4N}\delta_{il,jk}2\Lambda_{q}2\mathcal{O}_{il}^{\dag}\mathcal{O}_{il},\\
				\overline{g_{q}\Phi_{il1}^{-q}\Phi_{jk1}^{q}}
				&=\overline{g_{q}\Phi_{il2}^{-q}\Phi_{jk2}^{q}}=\frac{J\Lambda_{q}}{N}\delta_{il,jk},\\
				\overline{\Phi_{il1}^{-q}\Phi_{jk2}^{q}}&=0,
			\end{aligned}
		\end{equation}
		where $J>0$.
		Thus $\Phi_{il1}^{-q}$ is orthogonal with $\Phi_{jk2}^{q}$
		(as long as the $\Phi_{il}$ is approximately treated as independent of $\Phi_{jk}$ in the small $\Lambda_{q}$ limit (e.g., the  limit)).
		Combined with Eq.(\ref{Chi}), we know the variance ${\rm Var}(g_{q}\Phi_{il1}^{-q}\Phi_{il1}^{q})={\rm Var}(g_{q}\Phi_{jk2}^{-q}\Phi_{jk2}^{q})=0$,
		which is different to the result of next section.
		
		In the small (but finite) $\Lambda_{q}$ limit, according to semicircle law,
		the spectral function of single fermion reads
		\begin{equation} 
			\begin{aligned}
				\rho(\varepsilon)=\frac{1}{\pi J}\sqrt{1-\frac{\varepsilon^{2}}{4J^{2}}}
				=\frac{1}{2\pi J^{2}}\sqrt{4J^{2}-\varepsilon^{2}},
			\end{aligned}
		\end{equation}
		where 
		\begin{equation} 
			\begin{aligned}
				J^{2}=\overline{(g_{q}\Phi^{-q}_{1il}\Phi^{q}_{2jk})^{2}}\frac{N^{2}}{\Lambda_{q}^{2}}.
			\end{aligned}
		\end{equation}
		The mean value of eigenvalues is thus
		\begin{equation} 
			\begin{aligned}
				|\int_{\varepsilon<0}d\varepsilon \varepsilon\rho(\varepsilon)|\approx \frac{4J}{3\pi}.
			\end{aligned}
		\end{equation}
		Then we obtain that the matrices $g_{q}\Phi^{-q}_{1il}\Phi^{q}_{2jk}$
		and $\Phi^{-q}_{1il}$ and $\Phi^{q}_{2jk}$ are 
		$\frac{N^{2}}{\Lambda_{q}}\times\frac{N^{2}}{\Lambda_{q}}$ (a $\infty\times\infty$) matrix,
		and now these matrices are automatically diagonalized.
		In such a configration constructed by us,
		$g_{q}\Phi^{-q}_{1il}\Phi^{q}_{2jk}$ can not be simply viewed as a product of matrices
		$g_{q}\Phi_{il\sigma}\Phi_{jk\sigma'}$ and $g_{q}\Phi^{-q}\Phi^{q}$,
		since  $g_{q}\Phi_{il\sigma}\Phi_{jk\sigma'}$ is a $N^{2}\times N^{2}$ matrix
		while $g_{q}\Phi^{-q}\Phi^{q}$ is a $\Lambda_{q}^{-1}\times \Lambda_{q}^{-1}$ diagonal matrix
		($q\neq 0$ here).
		Instead, it requires mapping $il(jk)\rightarrow \frac{il(jk)}{\sqrt{\Lambda_{q}}}$
		and $q\rightarrow q\frac{1}{N^{2}}$ (to realizes $\delta\Lambda_{q}'=\Lambda_{q}^{'2}=\Lambda_{q}^{2}\frac{1}{N^{2}}$).
		This is the  phase with gapless  mode,
		and it requires $\Lambda_{q}^{-1}\gg N$.

		\section{Remove the correlation between $\Phi_{il}$ and $\Phi_{jk}$ by summing over $q$}
		
		The  phase can be gapped out
		due to the broken symmetry by finite eigenvalue of $\overline{(g_{q}\Phi_{1il}\Phi_{2jk})^{2}}$ (or $\overline{(g_{q}\Phi_{il\sigma}\Phi_{jk\sigma'})^{2}}$).
		To understand this,
		it is more convenient to use another configuration,
		where we carry out the summation over $q$ first in Eq.(34),
		instead of carrying out the disorder averages over $ijkl$ and $q$ in the same time.
		Then the disorder average over fermion indices $i,j,k,l$ simply results in
		\begin{equation} 
			\begin{aligned}
				\overline{(g_{q}\Phi_{1il}\Phi_{2jk})^{2}}
				=\frac{J_{il;jk}^{'2}}{4N^{2}},
			\end{aligned}
		\end{equation}
		which can be calculated as
		\begin{equation} 
			\begin{aligned}
				\overline{(g_{q}\Phi_{1il}\Phi_{2jk})^{2}}
				=&\overline{(\sum_{q}^{\Lambda_{q}^{-1}}g_{q}\Phi^{-q}_{1il}\Phi^{q}_{1jk})^{2}}\\
				=&\overline{(g_{q}\Phi^{-\delta\Lambda_{q}}_{1il}\Phi^{\delta\Lambda_{q}}_{1jk}
					+g_{q}\Phi^{-2\delta\Lambda_{q}}_{1il}\Phi^{2\delta\Lambda_{q}}_{1jk}+\cdot\cdot\cdot)^{2}}\\
				=&\overline{(g_{q}\Phi^{-\delta\Lambda_{q}}_{1il}\Phi^{\delta\Lambda_{q}}_{1jk})^{2}
					+(g_{q}\Phi^{-2\delta\Lambda_{q}}_{1il}\Phi^{2}_{1jk})^{2\delta\Lambda_{q}}+\cdot\cdot\cdot}\\
				=&\frac{J^{2}}{4N^{2}}\sum_{q}(\Phi^{-q}\Phi^{q})^{2}
				=\frac{J^{2}4\Lambda_{q}}{4N^{2}}
				=\frac{J^{2}\Lambda_{q}}{N^{2}},
			\end{aligned}
		\end{equation}
		i.e., $J_{il;jk}^{'2}=4\Lambda_{q}J_{il;jk}^{2}$.
		The third line is due to the fact about variance of Gaussian variables:
		${\rm Var}(A+B)={\rm Var}\ A+{\rm Var}\ B$ where $A$ and $B$ are independent with each other.
		The fourth line is because 
		\begin{equation} 
			\begin{aligned}
				{\rm Var}(\Phi^{q})=&\overline{(\Phi^{q})^{2}}-(\overline{\Phi^{q}})^{2}=2\Lambda_{q}-0=2\Lambda_{q},\\
				{\rm Var}(\Phi^{-q}\Phi^{q})=&\overline{(\Phi^{-q}\Phi^{q})^{2}}-(\overline{\Phi^{-q}\Phi^{q}})^{2}=4\Lambda^{2}_{q}-(2\Lambda_{q})^{2}=0,\\
				\sum_{q}\Phi^{-q}\Phi^{q}=&\overline{\Phi^{-q}\Phi^{q}}\Lambda_{q}^{-1}=2,\\
				\sum_{q}(\Phi^{-q}\Phi^{q})^{2}=&\overline{(\Phi^{-q}\Phi^{q})^{2}}\Lambda_{q}^{-1}=4\Lambda_{q},
			\end{aligned}
		\end{equation}
		where $\Phi^{-q}\Phi^{q}$ is obviously not a Gaussian variable
		(just like the $\Phi_{il}\Phi_{jk}$ except in the  limit),
		and 
		$\overline{\Phi^{-q}\Phi^{q}}=\overline{(\Phi^{q})^{2}}\neq 
		\overline{\Phi^{-q}}\overline{\Phi^{q}}\neq 0$ because it is impossible to make $\Phi^{-q}$ and $\Phi^{q}$ orthogonal 
		to each other.
		That is to say, although $\Phi^{-q}$ and $\Phi^{q}$ are Gaussian variables with zero mean $\overline{\Phi^{-q}}=\overline{\Phi^{q}}=0$,
		their product $\Phi^{-q}\Phi^{q}$ is not a Gaussian variable and do not have zero mean.
		This is different to the variance of Chi-square variable $\Phi_{1il}\Phi_{2jk}$ which is finite due to the zero mean 
		$\overline{\Phi_{1il}\Phi_{2jk}}=\overline{\Phi_{1il}}\overline{\Phi_{2jk}}=0$,
		by treating them to be approximately mutrually orthogonal (i.e., $\Phi_{il}$ approximately independent with $\Phi_{jk}$).
		Here we note that following relations in new configuration
		\begin{equation} 
			\begin{aligned}
				\overline{\sqrt{|g_{q}|}\Phi_{1il}}=&\overline{\sqrt{g_{q}}\Phi_{2jk}}=0,\\
				{\rm Var}(\sqrt{|g_{q}|}\Phi_{1il})=&\overline{g_{q}\Phi_{1il}^{2}}=\frac{J}{2N}2=\frac{J}{N},\\
				{\rm Var}(\sqrt{|g_{q}|}\Phi_{2jk})=&\overline{g_{q}\Phi_{2jk}^{2}}=\frac{J}{2N}2=\frac{J}{N},\\
				{\rm Var}(g_{q}\Phi_{1il}\Phi_{1jk})
				=&\overline{(g_{q}\Phi_{1il}\Phi_{1jk})^{2}}-\overline{g_{q}\Phi_{1il}\Phi_{1jk}}^{2}\\
				=&\frac{J^{2}}{4N^{2}}4\Lambda_{q}\delta_{il,jk}-(\frac{J}{2N}2)^{2}\delta_{il;jk}\\
				=&\frac{J^{2}}{N^{2}}(\Lambda_{q}-1)\delta_{il,jk},\\
				\overline{g_{q}\Phi_{1il}\Phi_{2jk}}=&0.
			\end{aligned}
		\end{equation}
		
		Under this configuration,
		by approximately treating $\Phi_{1il}$ and $\Phi_{2jk}$ to be muturally orthogonal,
		they can be viewed as two vectors, and each of them contains $N^{2}$ components,
		then $(g_{q}\Phi_{1il}\Phi_{2jk})$ is a $N^{2}\times N^{2}$ matrix with complex eigenvalues.
		But note that, away from the $\Lambda^{-1}_{q}\rightarrow \infty$ limit, this construction fail because
		exactly speaking, $(g_{q}\Phi_{1il}\Phi_{2jk})$ (after summation over $q$) is a $N\times N\times N=N^{3}$ matrix
		(unlike the $N\times N$ $H_{\flat=2}$  or the $N\times N\times N\times N=N^{4}$ $H_{\flat=4}$ )
		due to the bound state property, i.e., over the fermion indices $i,j,k,l$,
		one of them is always identified by the other three, so there are at most three independent indices (degrees of freedom).
		
		A precondition to treat Gaussian variables $i\Phi_{il}$ and $\Phi_{jk}$ murtually orthogonal (independent),
		is that it must away from the $\Lambda_{q}^{-1}\rightarrow \infty$ limit,
		since too small sample number will makes the matrix $g_{q}\Phi_{-q}\Phi_{q}$ leaves away from the Gaussian distribution
		according to central limit theorem, and thus the disorder average over $q$ can not be successively carried out,
		that is why we instead make the summation over $q$.
		Then, the relation $\Lambda_{q}^{-1}<N^{2}/2$ ($\Lambda_{q}>2/N^{2}$) indicates the large number of $N$,
		which preserves the Gaussian distribution of $\Phi_{il}$ and $\Phi_{jk}$ and also makes the disorder average over fermion indices
		to matrix $g_{q}\Phi_{il\sigma}\Phi_{jk\sigma'}$ more reliable, despite the 
		indices $il$ and $jk$ are not completely independent but correlated by some certain mechanism before the summation over $q$.

		Then we turning to the matrix 
		\begin{equation} 
			\begin{aligned}
				g_{q}\Phi^{-q}_{il\sigma}\Phi^{q}_{jk\sigma'}=ig_{q}\Phi^{-q}_{1il}\Phi^{q}_{2jk}-ig_{q}\Phi^{-q}_{2il}\Phi^{q}_{1jk},
			\end{aligned}
		\end{equation}
		which is also a $N^{2}\times N^{2}$ matrix now and is Hermitian 
		(whose eigenvectors and eogenvalues are much more easy to be solved) with all diagonal elements be zero.
		In this scheme, to make sure $g_{q}\Phi_{il\sigma}\Phi_{jk\sigma'}$ is a $N^{2}\times N^{2}$ matrix,
		the disorder average over $i,j,k,l$ must be done after the summation over $q$.
		Then to diagonalizing the $N^{2}\times N^{2}$ matrix $g_{q}\Phi^{-q}_{il\sigma}\Phi^{q}_{jk\sigma'}$,
		it requires $\Lambda_{q}^{-1}<N^{2}/2$ to make sure all the vectors $\Phi_{1il}$ and $\Phi_{2jk}$ are orthogonal with each other
		within the matrix $g_{q}\Phi^{-q}_{il\sigma}\Phi^{q}_{jk\sigma'}$.
		This is because there at most exists $N^{2}$ vectors can orthogonal with each other in $N^{2}$-dimensional space
		(formed by $N^{2}$-component vectors).
		In other word,
		the propose of this is to make sure vectors
		$\Phi^{-\delta\Lambda_{q}}_{il},\Phi^{-2\delta\Lambda_{q}}_{il},\cdot\cdot\cdot
		\Phi_{il}^{-\Lambda_{q}},\Phi^{\delta\Lambda_{q}}_{jk},\Phi^{2\delta\Lambda_{q}}_{jk},\cdot\cdot\cdot\Phi_{jk}^{\Lambda_{q}}$ are 
		orthogonal to each other.
		
		Then there are $N^{2}-2\Lambda_{q}^{-1}$ eigenvectors with eigenvalues equal zero (correspond to the ground state),
		i.e., 
		\begin{equation} 
			\begin{aligned}
				{\rm Det}[g_{q}\Phi^{-q}_{il\sigma}\Phi^{q}_{jk\sigma'}]=0,
			\end{aligned}
		\end{equation}
		and $2\Lambda_{q}^{-1}$ eigenvectors $\Phi_{\sigma}=\frac{\Phi_{1}^{-q}\pm i\Phi_{2}^{q}}{\sqrt{2}}$ 
		with eigenvalue $\pm \frac{J}{2\Lambda_{q}^{-1/2}N}$.
		This can be verified by the rule that for Hermitian matrix  the eigenvectors corresponding to different eigenvalues are orthogonal to each other,
		\begin{equation} 
			\begin{aligned}
				(\frac{\Phi_{1}^{-q}+ i\Phi_{2}^{q}}{\sqrt{2}})^{H}\cdot \frac{\Phi_{1}^{-q}- i\Phi_{2}^{q}}{\sqrt{2}}
				=\frac{1}{2}(\Phi_{1}^{-q}\cdot\Phi_{1}^{-q}-i\Phi_{1}^{-q}\cdot\Phi_{2}^{q}-i\Phi^{-q}_{1}\cdot\Phi^{q}_{2}-\Phi_{2}^{q}\cdot\Phi_{2}^{q})
				=0,
			\end{aligned}
		\end{equation}
		where $\Phi^{-q}_{1}\cdot\Phi^{q}_{2}=\Phi_{1}^{-q}\cdot\Phi_{2}^{q}=0$ since they are orthogonal to each other,
		and superscript $H$ denotes the transpose conjugation (Hermitian conjugate).
		The result of Eq.(35) is used here.
		
		In the special case of $\Lambda_{q}^{-1}=N^{2}/2$,
		we have, in matrix $g_{q}\Phi_{il\sigma}\Phi_{jk\sigma'}$, 
		$N^{2}$ eigenvectors $\frac{\Phi_{1}^{-q}\pm i\Phi_{2}^{q}}{\sqrt{2}}$ with eigenvalue $\pm \frac{J\Lambda_{q}}{2\sqrt{2}}$.
		Then processing the disorder average over $ijkl$ to $g_{q}\Phi^{-q}_{\sigma}\Phi^{q}_{\sigma'}$,
		we have the variance 
		\begin{equation} 
			\begin{aligned}
				{\rm Var}(g_{q}\Phi^{-q}_{il\sigma}\Phi^{q}_{jk\sigma'})=
				\overline{(g_{q}\Phi^{-q}_{il\sigma}\Phi^{q}_{jk\sigma'})^{2}}=\overline{(ig_{q}\Phi^{-q}_{1il}\Phi^{q}_{2jk}-ig_{q}\Phi^{-q}_{2il}\Phi^{q}_{1jk})^{2}}
				=\frac{1}{4}\frac{J^{2}\Lambda_{q}}{N^{2}}=\frac{J^{2}}{2N^{4}},
			\end{aligned}
		\end{equation}
		since $\overline{g_{q}\Phi^{-q}_{il\sigma}\Phi^{q}_{jk\sigma'}}=0$.
		The factor $1/4$ origin from the spin degrees of freedom,
		and can be verified by the square of eigenvalue $\lambda^{2}$.
		Note that here the overline denotes only the disorder average over $ijkl$ index.
		This result is in consistent with the property of Wigner matrix in GUE 
		\begin{equation} 
			\begin{aligned}
				\overline{\lambda^{2}}=\overline{(g_{q}\Phi^{-q}_{il\sigma}\Phi^{q}_{jk\sigma'})^{2}}=\frac{J^{2}\Lambda_{q}}{4N^{2}}\sim O(N^{-2}),
			\end{aligned}
		\end{equation}
		in contracst with that in Gaussian orthogonal ensemble (GOE) which reads $O(N^{-2})(1+\delta_{il,jk})$.
		Here $\lambda$ denote the eigenvalues.
		The GUE with $\sum \lambda^{2}=\frac{J^{2}\Lambda_{q}}{4}$ thus corresponds to the  non-fermi liquid case,
		with continuous distributed peaks in the  fermion spectral function,
		i.e., the level statistics agree with the GUE distribution, and the set of eigenvalues follow an ascending order.
		In GUE, we also have the relation
		\begin{equation} 
			\begin{aligned}
				\langle \mathcal{O}_{il}\mathcal{O}_{jk}^{\dag}\rangle
				=\langle c^{\dag}_{i\sigma}c_{l\sigma}c^{\dag}_{j\sigma'}c_{k\sigma'}\rangle
				=\langle c^{\dag}_{i\sigma}c^{\dag}_{j\sigma'}c_{k\sigma'}c_{l\sigma}\rangle
				=\langle c^{\dag}_{i\sigma}c_{l\sigma}\rangle\langle c^{\dag}_{j\sigma'}c_{k\sigma'}\rangle-
				\langle c^{\dag}_{i\sigma}c_{k\sigma'}\rangle\langle c^{\dag}_{j\sigma'}c_{l\sigma}\rangle,
			\end{aligned}
		\end{equation}
		at zero temperature.
		While the GOE correponds to the case of nonzero pairing order parameter 
		(in which case pair condensation happen at temperature lower than the critical one),
		and thus admit the anomalous terms.
		In GOE we have
		\begin{equation} 
			\begin{aligned}
				\overline{\lambda^{2}}=&\overline{(g_{q}\Phi^{-q}_{il\sigma}\Phi^{q}_{jk\sigma'})^{2}}=\frac{J^{2}\Lambda_{q}}{4N^{2}}+\frac{O(\delta_{il,jk})}{N^{2}},\\
				\sum\lambda^{2}=&\frac{J^{2}\Lambda_{q}}{4}+O(\delta_{il,jk}).
			\end{aligned}
		\end{equation}
		Thus the GOE has a level repulsion slightly larger than that of GUE in the small level spacing limit during the level statistic.
		Then if we turn to the many-body localized phase where the thermalization (chaotic) is being suppressed by the stronger disorder,
		the level statistic follows the Poisson distribution.
		Since $g\Phi_{il\sigma}\Phi_{jk\sigma'}$ is not a positive-define matrix,
		the largest eigenvalues splitting happen which corresponds to the discrete spectrum with the level statistics agree with Poisson distribution,
		i.e., it has the largest eigenvalues $\lambda_{max}=\pm \frac{J\lambda_{q}^{1/2}}{2\sqrt{2}}$ and $\frac{N^{2}-2}{2}$ eigenvalues $\frac{1}{N^{2}-2}$
		and $\frac{N^{2}-2}{2}$ eigenvalues $\frac{-1}{N^{2}-2}$.
		Such a distribution of eigenvalues implies the existence of off-diagonal long range order.
		When the pair condensation happen,
		the above relation becomes $\langle c^{\dag}_{i\sigma}c^{\dag}_{j\sigma'}\rangle\langle c_{k\sigma'}c_{l\sigma}\rangle$,
		with the pairing order parameter 
		\begin{equation} 
			\begin{aligned}
				\Delta_{0}=\langle c_{k\sigma'}c_{l\sigma}\rangle
				=\frac{\sum_{kl}\sum_{\sigma\sigma'} c_{k\sigma'}c_{l\sigma}}{2N^{1/2}}
			\end{aligned}
		\end{equation}
		where the factor $(2N^{1/2})^{-1}$ origin from the result of disorder average
		\begin{equation} 
			\begin{aligned}
				\overline{\Phi_{kl}^{2}}=\frac{J}{2N}2=\frac{J}{N},\ 
				\overline{\Phi_{kl,\sigma\sigma'}^{2}}=\frac{J}{4N}.
			\end{aligned}
		\end{equation}
		The $N^{2}\times N^{2}$ positive-define matrix $\langle c^{\dag}_{i\sigma}c^{\dag}_{j\sigma'}\rangle\langle c_{k\sigma'}c_{l\sigma}\rangle$
		has summation of eigenvalues corresponds to the total number of pairs $\sum \lambda\le N^{2}$ and thus $\overline{\lambda}\le 1$,
		i.e., $\lambda_{max}=N^{2}=Sp |c^{\dag}_{i\sigma}c^{\dag}_{j\sigma'}\rangle\langle c_{k\sigma'}c_{l\sigma}|\equiv Sp b_{ij}^{\dag}b_{kl}$.
		We also found that, once the boson-fermion interacting term is taken into account,
		the maximum eigenvalue reduced to:
		For $ij\neq kl$,
		$\lambda_{max}= {\rm Sp} b_{ij}^{\dag}c^{\dag}cb_{kl}\le {\rm Sp} (b_{ij}^{\dag}c^{\dag}cb_{kl}+cb_{kl}b_{ij}^{\dag}c^{\dag})=0$;
		For $ij=kl$ ($\Lambda_{q}=0$),
		$\lambda_{max}= {\rm Sp} b_{ij}^{\dag}c^{\dag}cb_{kl}\le {\rm Sp} (b_{ij}^{\dag}c^{\dag}cb_{kl}+cb_{kl}b_{ij}^{\dag}c^{\dag})={\rm Sp} cc^{\dag}
		={\rm Sp}(1-c^{\dag}c)=1-{\rm Sp} c^{\dag}c\le 1$.
		The superconductivity emerge when $\Delta_{0}$ condenses,
		and in large -N limit,
		the renormalized Green's function reads
		\begin{equation} 
			\begin{aligned}
				G(i\omega)'=\frac{G(i\omega)}{1+J^{2}|\Delta_{0}^{2}|G^{2}(i\omega)}.
			\end{aligned}
		\end{equation}
		For a further study about this renormalization effect, see Ref.\cite{Patel A A,Pikulin D I}.

		In this case, the coupling $J$ within spectral function reads 
		\begin{equation} 
			\begin{aligned}
				J^{2}=2N^{4}\overline{(g_{q}\Phi^{-q}_{il\sigma}\Phi^{q}_{jk\sigma'})^{2}},
			\end{aligned}
		\end{equation}
		In the $\Lambda_{q}^{-1}\ll N^{2}/2$ limit, we can easily know that $J$ is vanishingly small,
		and the bound stateic dynamic then dominates over the  dynamic,
		and the system exhibits fermi liquid feature.
		While for $0\ll \Lambda_{q}^{-1}<N^{2}/2$, the system exhibits disordered fermi liquid feature with sharp Landau quasiparticles,
		and for positive define matrix $g_{q}\Phi_{il\sigma}\Phi_{jk\sigma'}$,
		since every zero eigenvalue corresponds to a ground state, there are $N^{2}-2\Lambda_{q}^{-1}$ ground states,
		and thus the system exhibits degeneracy $2^{N^{2}-2\Lambda_{q}^{-1}}$.
		While in the case of $\mu\gg g_{q}$, the billinear term as a disorder will gap out the system and lift the degeneracy in ground state,
		although in some certain systems\cite{Graser S} the near nesting of fermi surface sheets can prevent the increase of degeneracy by disorder.
		Here the bilinear term is absent but the finite value of variance ${\rm Var}(g_{q}\Phi_{il\sigma}\Phi_{jk\sigma'})$ with $\Lambda_{q}^{-1}<N^{2}/2$
		plays its role and drives the  non-fermi liquid state toward the disordered fermi liquid ground state.
		
		Finally, we conclude that
		in $\Lambda_{q}^{-1}\ll N^{2}/2$ case,
		although $g_{q}\Phi_{il\sigma}\Phi_{jk\sigma'}$ is a Hermitian matrix with randomly independent elements and large $N$,
		and each of its matrix elements follows the same distribution (distribution of Chi-square variables),
		the eigenvalue distribution does not follows the semicircle law.
		This is because, for $\Lambda_{q}^{-1}< N^{2}/2$, $N^{2}$-component vectors $\Phi^{-q}_{1},\Phi^{q}_{2}$ are mutually orthogonal,
		which leads to large degeneracy in ground state.
		In this case, the spectral function does not follows the semicircle law, but exhibits three broadened peaks locate on
		the energies $\varepsilon=0,\pm \frac{J\Lambda_{q}^{1/2}}{2N}$, with heights correspond to the numbers of the corresponding eigenvectors.

		\section{$H_{\flat=2}$  bound state model in $\Lambda_{q}=0$}
		
		Base on the above discussions, we further extend the $\Lambda_{q}^{-1}\gg N$ limit
		to $\Lambda_{q}=0$ limit, i.e., the $H_{\flat=2}$  model.
		This is the limit that does not allows any bound state to exist,
		and now $\Phi_{il}, \Phi_{jk}$ are no longer Gaussian variables
		since $\phi_{i}$ can be treated the same as $\phi_{l}$,
		but $\Phi_{il}\Phi_{jk}$ (and $\Phi_{ik}$ as defined below) becomes a Gaussian variable,
		and
		the matrix $g_{q}\Phi_{il\sigma}\Phi_{jk\sigma'}$ follows Gaussian distribution.
		The Hamiltonian can now be reconstructed as
		\begin{equation} 
			\begin{aligned}
				H_{p}
				=&-g_{q}\sum_{ijkl}^{N}\Phi_{ik}\Phi_{jl}c^{\dag}_{i\sigma}c_{k\sigma'}c^{\dag}_{j\sigma'}c_{l\sigma}\\
				=&-g_{q}\sum_{ijkl}^{N}\Phi_{ik}c^{\dag}_{i\sigma}c_{k\sigma'}(\Phi_{ik}c^{\dag}_{i\sigma}c_{k\sigma'})^{*}\\
				=&-g_{q}\sum_{ik}^{N}(\Phi_{ik}c^{\dag}_{i\sigma}c_{k\sigma'})^{2},
			\end{aligned}
		\end{equation}
		where $\Phi_{ik}^{*}=\Phi_{jl}$, and $\overline{\Phi_{ik}}=0,\overline{\Phi_{jl}}=0$.
		For the special case that $i=k$, it becomes the Wishart- model\cite{Iyoda E}
		which is integrable.
		Then since $g_{p}<0$ for attractive bound state, we obtain that the eigenvalue of $H_{p}$ is positive.
		This guarantees the Gaussian distributions of $\sqrt{|g_{q}|}\Phi_{ik}$ and $\sqrt{|g_{q}|}\Phi_{jl}$,
		have the deviation reads
		\begin{equation} 
			\begin{aligned}
				\sum_{i>k}\overline{g_{q}\Phi_{ik}^{2}}&=\sum_{j>l}\overline{g_{q}\Phi_{jl}^{2}}
				=\overline{g_{q}\Phi_{ik}^{2}}\frac{N(N-1)}{2}\\
				=&\frac{J}{2(N^{1/4})^{2}((N-1)^{1/4})^{2}}\frac{N(N-1)}{2}
				=\frac{J}{2N^{1/2}(N-1)^{1/2}}\frac{N(N-1)}{2}\\
				\approx & \frac{J}{2N}\frac{N(N-1)}{2}
				=\frac{J(N-1)}{4}.
			\end{aligned}
		\end{equation}
		Then the Euclidean time path integral reads
		\begin{equation} 
			\begin{aligned}
				Z=&\int D[c^{\dag},c]e^{-S},\\
				S_{0}=&\int d\tau d\tau'[\sum_{i}^{N}c^{\dag}_{i}(\tau)\partial_{\tau}(\tau-\tau')c_{i}(\tau')+\sum_{k}^{N-1}c^{\dag}_{k}(\tau)\partial_{\tau}(\tau-\tau')c_{k}(\tau')],\\
				S_{int}=&\int d\tau d\tau'
				[-g_{q}\sum_{ik}\Phi_{ik}c^{\dag}_{i\sigma}c_{k\sigma'}(\Phi_{ik}c^{\dag}_{i\sigma}c_{k\sigma'})^{*}].
			\end{aligned}
		\end{equation}
		Using the replica trick $\overline{{\rm ln}Z}=\lim_{n\rightarrow 0}\frac{\overline{Z^{n}}-1}{n}$ with the Gaussian average, we have
		\begin{equation} 
			\begin{aligned}
				Z=&
				\int D[c^{\dag},c]e^{-\int d\tau d\tau'[\sum_{i}^{N}c^{\dag}_{i}(\tau)\partial_{\tau}\delta(\tau-\tau')c_{i}(\tau)
					+\sum_{k}^{N-1}c^{\dag}_{k}(\tau)\partial_{\tau}\delta(\tau-\tau')c_{k}(\tau)-g_{q}^{-1}\psi_{1}(\tau)\psi_{2}(\tau')]}\\
				&\int D[\Phi_{ik},\Phi_{ik}^{*}]
				e^{-\int d\tau d\tau'(i\psi_{1}\Phi_{ik}c_{i}^{\dag}c_{k}+i\psi_{2}\Phi_{ik}^{*}c_{k}^{\dag}c_{i}-H.c.)}\\
				&\int D[\Phi_{ik},\Phi_{ik}^{*}] e^{-\frac{|\Phi_{ik}|^{2}}{\sigma^{2}}}e^{-\frac{|\Phi_{ik}^{*}|^{2}}{\sigma^{2}}}\\
				=&
				\int D[c^{\dag},c]e^{-\int d\tau d\tau'[\sum_{i}^{N}c^{\dag}_{i}(\tau)\partial_{\tau}\delta(\tau-\tau')c_{i}(\tau)
					+\sum_{k}^{N-1}c^{\dag}_{k}(\tau)\partial_{\tau}\delta(\tau-\tau')c_{k}(\tau)-g_{q}^{-1}\psi_{1}(\tau)\psi_{2}(\tau')]}\\
				&\int D[\Phi_{ik},\Phi_{ik}^{*}]
				e^{-\int d\tau d\tau'
					(i\psi_{1}\Phi_{ik}c_{i}^{\dag}c_{k}+i\psi_{2}\Phi_{ik}^{*}c_{k}^{\dag}c_{i}
					-i\psi_{1}^{*}\Phi_{ik}^{*}c_{k}^{\dag}c_{i}-i\psi_{2}^{*}\Phi_{ik}c_{i}^{\dag}c_{k})}\\
				&\int D[\Phi_{ik},\Phi_{ik}^{*}] e^{-\frac{|\Phi_{ik}|^{2}}{\sigma^{2}}}e^{-\frac{|\Phi_{ik}^{*}|^{2}}{\sigma^{2}}},
			\end{aligned}
		\end{equation}
		where $\sigma^{2}=\overline{\Phi_{ik}^{2}}=\frac{J}{2N^{1/2}(N-1)^{1/2}}\approx  \frac{J}{2N}$ is the variance of Gaussian variable $\Phi_{ik}$.
		
		Since 
		\begin{equation} 
			\begin{aligned}
				&\int d\Phi_{ik}d\Phi_{ik}^{*} e^{-\frac{|\Phi_{ik}|^{2}}{\sigma^{2}}}
				e^{-(i\psi_{1}\Phi_{ik}c^{\dag}_{i}c_{k}+H.c.)}\\
				&=\int d\Phi_{ik}d\Phi_{ik}^{*} e^{-\frac{|\Phi_{ik}|^{2}}{\sigma^{2}}}
				e^{-(i\psi_{1}\Phi_{ik}c^{\dag}_{i}c_{k}-i\psi_{1}^{*}\Phi_{ik}^{*}c^{\dag}_{k}c_{i})}\\
				&=\int d\Phi_{il}d\Phi_{il}^{*} e^{\frac{-1}{\sigma^{2}}
					(\Phi_{ik}-i\sigma^{2}\psi_{1}^{*}c^{\dag}_{k}c_{i})(\Phi_{ik}^{*}+i\sigma^{2}\psi_{1}c^{\dag}_{i}c_{k})}
				e^{\sigma^{2}(\psi^{*}_{1}(\tau)\psi_{1}(\tau)
					c_{k}^{\dag}(\tau)c_{i}(\tau)c_{i}^{\dag}(\tau')c_{k}(\tau')}\\
				&=e^{\sigma^{2}(\psi^{*}_{1}(\tau)\psi_{1}(\tau)
					c_{k}^{\dag}(\tau)c_{i}(\tau)c_{i}^{\dag}(\tau')c_{k}(\tau')},
			\end{aligned}
		\end{equation}
		the action can be written as
		\begin{equation} 
			\begin{aligned}
				S=&\int d\tau d\tau'
				[\sum_{i}c^{\dag}_{i}(\tau)\partial_{\tau}\delta(\tau-\tau')c_{i}(\tau)+\sum_{k}c^{\dag}_{k}(\tau)\partial_{\tau}\delta(\tau-\tau')c_{k}(\tau)
				-g_{q}^{-1}\psi_{1}(\tau)\psi_{2}^{*}(\tau')-g_{q}^{-1}\psi_{2}(\tau)\psi_{1}^{*}(\tau')]\\
				&-\int d\tau d\tau' \sigma^{2}(\psi^{*}_{1}(\tau)\psi_{1}(\tau)c_{k}^{\dag}(\tau)c_{i}(\tau)c_{i}^{\dag}(\tau')c_{k}(\tau')
				-\int d\tau d\tau' \sigma^{2}(\psi^{*}_{1}(\tau)\psi_{1}(\tau)c_{k}^{\dag}(\tau)c_{i}(\tau)c_{i}^{\dag}(\tau')c_{k}(\tau')\\
				=&\int d\tau d\tau'
				[\sum_{i}c^{\dag}_{i}(\tau)\partial_{\tau}\delta(\tau-\tau')c_{i}(\tau)+\sum_{k}c^{\dag}_{k}(\tau)\partial_{\tau}\delta(\tau-\tau')c_{k}(\tau)
				-g_{q}^{-1}\psi_{1}(\tau)\psi_{2}^{*}(\tau')-g_{q}^{-1}\psi_{2}(\tau)\psi^{*}_{1}(\tau')]\\
				&-\int d\tau d\tau'\frac{(N-1)J}{2}\psi^{*}_{1}(\tau)\psi_{1}(\tau)G_{f}^{2}(\tau,\tau')
				-\int d\tau d\tau'\frac{(N-1)J}{2}\psi^{*}_{2}(\tau)\psi_{2}(\tau)G_{f}^{2}(\tau,\tau').
			\end{aligned}
		\end{equation}
		Defining $G_{f}(\tau,\tau')=N^{-1}\sum_{i}c^{\dag}_{i}(\tau')c_{i}(\tau)$ in mean field treatment,
		and using the identity
		\begin{equation} 
			\begin{aligned}
				1=\int D\Sigma DG_{f}(N\int d\tau d\tau'\Sigma(\tau,\tau')G_{f}(\tau',\tau)-\frac{1}{N}\sum_{i}c_{i}^{\dag}(\tau')c_{i}(\tau)),
			\end{aligned}
		\end{equation}
		we obtain
		\begin{equation} 
			\begin{aligned}
				S
				=&-(2N-1)\int d\tau d\tau'{\rm Tr}{\rm ln}[\partial_{\tau}+\Sigma(\tau,\tau')]
				-\int d\tau d\tau'{\rm Tr}{\rm ln}
				\begin{pmatrix}
					\psi_{1} & \psi_{2}
				\end{pmatrix}
				\begin{pmatrix}
					0 & -g_{q}^{-1}\\
					-g_{q}^{-1} & 0
				\end{pmatrix}
				\begin{pmatrix}
					\psi_{1}^{*} \\
					\psi_{2}^{*}
				\end{pmatrix}\\
				&-\int d\tau d\tau'\frac{(N-1)J}{2}\psi^{*}_{1}(\tau)\psi_{1}(\tau')G_{f}^{2}(\tau,\tau')
				-\int d\tau d\tau'\frac{(N-1)J}{2}\psi^{*}_{2}(\tau)\psi_{2}(\tau')G_{f}^{2}(\tau,\tau')\\
				&+(2N-1)\Sigma(\tau,\tau')G_{f}(\tau',\tau),
			\end{aligned}
		\end{equation}
		thus
		\begin{equation} 
			\begin{aligned}
				\frac{\partial S}{\partial \Sigma(\tau',\tau)}=&(2N-1)G_{f}(\tau,\tau')-\frac{2N-1}{-i\omega+\Sigma(\tau,\tau')},\\
				\frac{\partial S}{\partial G_{f}(\tau',\tau)}=&(2N-1)\Sigma(\tau,\tau')-\frac{J(N-1)}{2}\psi^{*}_{1}(\tau)\psi_{1}(\tau')G_{f}(\tau',\tau)
				-\frac{J(N-1)}{2}\psi^{*}_{2}(\tau)\psi_{2}(\tau')G_{f}(\tau',\tau).
			\end{aligned}
		\end{equation}

		In the above equation Eq.(53),
		we use the relation
		\begin{equation} 
			\begin{aligned}
				e^{g_{q}\Phi_{ik}\Phi_{ik}^{*}}
				=&\int D[\psi]e^{g_{q}^{-1}\psi_{1}\psi_{2}-i\psi_{1}\Phi_{ik}-i\psi_{2}\Phi_{ik}^{*}},
			\end{aligned}
		\end{equation}
		since
		\begin{equation} 
			\begin{aligned}
				\int D[\psi]e^{g_{q}^{-1}(\psi-i\Phi g_{q})^{2}}=1.
			\end{aligned}
		\end{equation}
		Then according to another identity
		\begin{equation} 
			\begin{aligned}
				\int D[\psi]e^{-\int d\tau d\tau'(-ig_{q}\psi_{1}(\tau)\psi_{2}(\tau'))}=e^{{\rm Tr}{\rm ln}(-ig_{q})}=1,
			\end{aligned}
		\end{equation}
		we have
		\begin{equation} 
			\begin{aligned}
				\int d\tau d\tau'(ig_{q}\psi_{1}(\tau)\psi_{2}(\tau'))=\int d\tau d\tau' g_{q}^{-1}(\psi_{1}(\tau)-i\Phi_{ik}^{*}(\tau')g_{q})(\psi_{2}(\tau')-i\Phi_{ik}(\tau)g_{q}).
			\end{aligned}
		\end{equation}
		
		In this case ($\Lambda_{q}=0$), $|g_{q}|\equiv g\rightarrow \infty$ which corresponds to the tightly bounded particles,
		and the bound state system is in ground state with zero energy,
		and the zero eigenstate (of matrix element $\Phi_{ik}$) corresponds to zero density-of-states
		since
		\begin{equation} 
			\begin{aligned}
				D(\varepsilon=0)=\rho(\varepsilon=0)=2^{N/2}\frac{1}{\sqrt{2\pi}\sqrt{\sum_{i>k}|\Phi_{ik}|^{2}}}=2^{N/2}\sqrt{\frac{2}{J(N-1)\pi}}
				\rightarrow 0 \ (J\rightarrow \infty),
			\end{aligned}
		\end{equation}
		where $2^{N/2}$ is the number of states for $N$ pair of fermions (or $N$ bound state modes; each mode comtains four fermions
		and $N/2$ corresponds to number of creation operators).
		Here the eigenstates follows the Gaussian distribution.
		This is different with the $g_{q}>0$ (repulsive bound state) case,
		in which case the eigenvalues are negative and follows the semicircle law distribution.
		In repulsive bound state case,
		the mean value of maximum eigenvalue (a summation over all $2^{N/2}$ states) of $H_{p}$ is 
		$(\frac{4J}{3\pi})^{2}$ according to Eq.(39),
		where $J=2N\overline{\Phi_{ik}^{2}}$.
		To obtain the ground state entropy,
		we firstly write the partition function as
		\begin{equation} 
			\begin{aligned}
				Z=&{\rm Tr}e^{-\beta H_{p}}=\int d\varepsilon \rho(\varepsilon)e^{-\beta\varepsilon^2}
				=\int^{\infty}_{0} d\varepsilon 2^{N/2}\sqrt{\frac{2}{J(N-1)\pi}}e^{-\frac{\varepsilon^{2}}{2\frac{J(N-1)}{4}}}
				e^{-\beta\varepsilon^{2}}
				=\frac{2^{-1+(1+N)/2}\sqrt{\frac{1}{J(N-1)}}}{\beta+\frac{2}{J(N-1)}},
			\end{aligned}
		\end{equation}
		then the entropy density reads
		\begin{equation} 
			\begin{aligned}
				S
				=&\frac{1}{N}(\frac{\partial \overline{\varepsilon^{2}}}{\partial T}+\frac{-\partial F}{\partial T})
				=\frac{1}{N}({\rm ln}Z+T\frac{\partial {\rm ln}Z}{\partial T})
				=\frac{1}{N}({\rm ln}Z-\beta\frac{\partial {\rm ln}Z}{\partial \beta})\\
				=&\frac{\beta 2^{-1+(-1-N)/2+(1+N)/2}}{(\beta+\frac{2}{J(N-1)})N}+{\rm ln}\frac{2^{-1+(1+N)/2}\sqrt{\frac{1}{J(N-1)}}}{\sqrt{\beta+\frac{2}{J(N-1)}}},
			\end{aligned}
		\end{equation}
		where $\overline{\varepsilon^{2}}=\frac{\sum_{\varepsilon}\varepsilon^{2}e^{-\beta\varepsilon^{2}}}{Z}$ is the average energy of Haimiltonian $H_{p}$
		and $f=\frac{F}{N}=\frac{-T{\rm ln}Z}{N}$ is the free energy density,
		and we found 
		\begin{equation} 
			\begin{aligned}
				\lim_{N\rightarrow \infty}S=\frac{{\rm ln}2}{2}.
			\end{aligned}
		\end{equation}
		This nonzero ground state entropy does not related to the temperature,
		which is an important feature of  system.
		The specific heat can be obtained as
		\begin{equation} 
			\begin{aligned}
				C_{v}
				=&-\beta\frac{\partial S}{\partial \beta}
				=\frac{\beta^{2} 2^{-1+(-1-N)/2+(1+N)/2}}{(\beta+\frac{2}{J(N-1)})^{2}N}.
			\end{aligned}
		\end{equation}
		And the heat capacity can be obtained as
		\begin{equation} 
			\begin{aligned}
				C_{p}
				=&\frac{\partial f}{\partial T}
				=\frac{\partial F}{N\partial T}
				=-T\frac{\partial {\rm ln}Z}{N\partial T}
				=\beta\frac{\partial {\rm ln}Z}{N\partial \beta}\\
				=&-\frac{\beta 2^{-1+(-1-N)/2+(1+N)/2}}{(\beta+\frac{2}{J(N-1)})N}.
			\end{aligned}
		\end{equation}

		For $N\times N$ matrix $\Phi_{ik}$ with large $N$, it is a
		Gaussian Wigner matrix as its each matrix elements are randomly independent and 
		follows the Gaussian distribution, unike the above $N^{2}\times N^{2}$ matrix $\Phi_{il}\Phi_{jk}$
		whose matrix elements (Chi-square random variables) follow the Gamma distributions.
		But the eigenvalues of these two matrices convergent to the semicircle law which is 
		an asymptotically free (freely independent) 
		analogue of the central limit theorem (for scalar probability theory).

		\section{Conclusion}
		
	In terms of the relative momentum during scattering between impurity and majority particles (or holes) $q$, the bound state forms at the pole of scattering amplitude where $q=ia^{-1}$, where $a$ is the scattering length,
		and the quantity $qa$ becomes an important characterisic scale in predicting the many-body behaviors.
		Note that we works on the condition that close to the half-filling
		with cutoff $\Lambda_{q}^{-1}\rightarrow \infty$ to satisfy the requirement of ensemble behavior.
		In such non-fermi liquid phase, since the $q$ is small, and the coupling is strong,
		the bound state is not well-defined unlike in the fermi liquid phase.
		In the opposite limit with $\Lambda_{q}^{-1}\ll N^{2}$,
		as we also discuss in this paper, the system exhibits (disordered) fermi liquid features,
		e.g., the sharp peaks in spectral function.
		While in the fermi liquid phase with well-defined fermi surface,
		the above characterisic scale should be replaced by $k_{F}a$ where $k_{F}$ is the fermi wave vector.
		
		In conclusion,
	although the bound state system described by a four-point interacting term is not a $H_{\flat=4}$  model,
		we can analyse the matrix in GUE by treating $g_{q}\Phi_{il\sigma}\Phi_{jk\sigma'}$ as a $N^{2}\times N^{2}$ matrix
		 after summation over $q$-superscript (i.e., the mean value times $\Lambda_{q}^{-1}$).
		In this case (as long as $\Lambda_{q}\neq 0$),
		the variables $\Phi_{il}$ and $\Phi_{jk}$ are Gaussian,
		while $g_{q}\Phi_{il\sigma}\Phi_{jk\sigma'}$ is a Chi-square variable.
		And in the mean time, since the $q$ component are summarized, the wave functions $\Phi_{il}$ and $\Phi_{jk}$ become completely independent.
		This is what we called $H_{\flat=2}\times H_{\flat=2}\times H_{\flat'=1}$ model,
		which can well describes the  dynamics of bound state system.
		Here the $\flat'=1$ is to combine the disorder average over $\Phi^{q}$ and $\Phi^{-q}$ components into the disorder average over 
		fermion indices.
		While in the long-wavelength limit ($\Lambda_{q}=0$),
		$\Phi_{il}$ and $\Phi_{jk}$ are no more Gaussian, 
		but the Hamiltonian can be reconstructed to contains the Gaussian variable $\Phi_{ik}$,
		where $\overline{\Phi_{ik}^{2}}=\overline{\Phi_{jl}^{2}}=\frac{J}{2N}$.
		Also, in this case, the  Hamiltonian (bound state no more exists now)
		is positive-define, which means the eigenstates corresponding to zero eigenvalue are the ground states,
		and all eigenvalues of matrix $g_{q}\Phi_{il\sigma}\Phi_{jk\sigma'}$ are positive.
		Lastly, we discuss the case of pair condensation,
		which happen at critical temperature, and leads to zero dressed (or undressed) boson self-energy.
		We deduce the free energy density and the divergent behaviors of anomalous propagator in this case.
		
		During the above basis transformation between the original bound state momentum basis and the  fermion indices basis,
		in the $\Lambda_{q}^{-1}\rightarrow\infty$ limit,
		only the statistical relations between $\phi_{i}$ and $\phi_{l}$ or $\phi_{j}$ and $\phi_{k}$ depends on the value of $\Lambda_{q}^{-1}$,
		but this dependence on $\Lambda_{q}^{-1}$ is also being replaced by number $N$ after the transformation.
		While the disorder average over $q$ is carried out seperately with that of $N$,
		which is allowed only under the condition $\Lambda_{q}^{-1}\gg N^{2}$,
		and in this case, the  physics cannot be realized if we do the summation over $q$ first
		which requires $\Lambda_{q}^{-1}<N^{2}/2$ to realize ensemble behavior.
		
		\clearpage
		
		\section{Appendix.A: Application of Hubbard-Stratonovich transformation and replica technique for coupled product}
			Using Baker-Campbell-Hausdorff formula, we can write a density matrix inclduing a coupled product $e^{x}e^{y}$ as
		\begin{equation} 
			\begin{aligned}
				{\rm ln}(e^{x}e^{y})=&
				\sum_{i_{1}\in Z^{*},j_{1}=0,1}\frac{[x^{(i_{1})},y^{(j_{1})}]}{(i_{1}+j_{1})i_{1}!j_{1}!}
				+\sum_{i_{1}\in Z^{*},j_{2}=0,1}
				\frac{[x^{(i_{1})},[y^{(j_{1})},[x^{(i_{2})},y^{(j_{2})}]]]}
				{(i_{1}+j_{1}+i_{2}+j_{2})i_{1}!j_{1}!i_{2}!j_{2}!}\\
				&=x+y+\sum_{i_{1}=1}^{\infty}\frac{[x^{(i_{1})},y]}{(i_{1}+1)i_{1}!}\\
				&
				+\sum_{i_{1}\ge 1}\frac{[x^{(i_{1})},y]}{(i_{1}+1)i_{1}!}
				+\sum_{i_{1}\ge 0,j_{1}\ge 1}
				\frac{[x^{(i_{1})},[y^{(j_{1})},x]]}{(i_{1}+j_{1}+1)i_{1}!j_{1}!}
				+\sum_{i_{1}\ge 0,j_{1}\ge 1,i_{2}\ge 1}
				\frac{[x^{(i_{1})},[y^{(j_{1})},[x^{(i_{2})},y]]}{(i_{1}+j_{1}+i_{2}+1)i_{1}!j_{1}!i_{2}!}
			\end{aligned}
		\end{equation}
		where
		$[x^{(2)},y]=[x,[x,y]]$ denotes the iterated Lie bracket.
		We consider only to the first order where
		\begin{equation} 
			\begin{aligned}
				{\rm ln}(e^{x}e^{y})=x+y+\frac{[x,y]}{2}.
			\end{aligned}
		\end{equation}
		Then we have
		\begin{equation} 
			\begin{aligned}
				&
				e^{x}e^{y}=e^{x+y+\frac{[x,y]}{2}},\\
				&
				e^{x}e^{y}e^{-x}=e^{y+[x,y]},\\
				&
				e^{-y}e^{x}e^{y}=e^{x+[x,y]},
			\end{aligned}
		\end{equation}
		which are consistent with inserting an operator identity
		through the Baker-Hausdorff lemma,
		\begin{equation} 
			\begin{aligned}
				&
				e^{x}e^{y}e^{-x}=
				e^{y}+[x,e^{y}]=e^{y}(1+\frac{[x,e^{y}]}{e^{y}})
				=e^{y}(1+[x,y])=e^{y}e^{[x,y]}=
				e^{y+[x,y]},\\
				&
				e^{-y}e^{x}e^{y}=e^{x}+[-y,e^{x}]
				=e^{x}(1+\frac{[-y,e^{x}]}{e^{x}})=e^{x}(1+[-y,x])
				=e^{x}e^{[-y,x]}
				=e^{x+[x,y]}.
			\end{aligned}
		\end{equation}

		Next we consider the Hamiltonian
		$h=e^{a}e^{b}$ where $[a,b]\neq 0$,
		the corresponding density matrix
		is $e^{-h}$.
		To perform the Hubbard-Stratonovich transformation,
		we introduce the identity
		\begin{equation} 
			\begin{aligned}
				\int \mathcal{D}[\psi] e^{-\psi_{1}\psi_{2}}=1.
			\end{aligned}
		\end{equation}
		
		In terms of the corresponding linear response 
		of effective Hamiltonian,
		we have another set of variable
		$\hat{\psi}_{1},\hat{\psi}_{2}$ that share the similar identity with
		$\psi_{1},\psi_{2}$:
		\begin{equation} 
			\begin{aligned}
				&		\hat{\psi}_{1}=\psi_{1}+\sum_{i}e^{a}_{i}x_{i},\\
				&		\hat{\psi}_{2}=\psi_{2}+\sum_{j}e^{b}_{j}x_{j},\\
			\end{aligned}
		\end{equation}
		where
		\begin{equation} 
			\begin{aligned}
				\int \mathcal{D}[\hat{\psi}] e^{-\hat{\psi}_{1}\hat{\psi}_{2}}=1,
			\end{aligned}
		\end{equation}
		implies the variable set $\psi_{1},\psi_{2}$ can be viewed as a saddle-point case of $\hat{\psi}_{1},\hat{\psi}_{2}$ which is specified by certain saddle-point coordinator-dependence
		(linearly independent of others),
		in terms of the fluctuational expansion around saddle-point.
		The latter terms describe the fluctuations,
		a set of coupled coefficients $x_{i},x_{j}$
		($[x_{i},x_{j}]\neq 0$) successfully decoupes the two sums including the taylor-expanded exponential terms
		$e^{a}$ and $e^{b}$, and in the mean time without change their product,
		\begin{equation} 
			\begin{aligned}
				\sum_{i}e^{a}_{i}x_{i}\sum_{j}e^{b}_{j}x_{j}
				=e^{a}e^{b}=e^{a+b+\frac{[a,b]}{2}}.
			\end{aligned}
		\end{equation}
From this, we can write the effective susceptibility as
\begin{equation} 
	\begin{aligned}
				\frac{\delta x_{j}}{\delta \phi_{i}}=
				\langle x_{i}x_{j}\rangle 
				=\frac{	\sum_{i}e^{a}_{i}x_{i}\sum_{j}e^{b}_{j}x_{j}}
				{	\sum_{i}e^{a}_{i}\sum_{j}e^{b}_{j}}
				=e^{\frac{[a,b]}{2}},
	\end{aligned}
\end{equation}
		where the corresponding symmetry property can be observed in terms of the zero functional derivatives of $x_{i}$ 
		with respect to the bosonic source field $\phi_{i}$.
		The corresponding linear response is
\begin{equation} 
	\begin{aligned}
		e^{-\beta H}\rightarrow e^{-\beta H}\mathcal{T}
		e^{\int d\tau x_{i}(\tau)\phi_{i'}(\tau)},
			\end{aligned}
\end{equation}
		where through bosonization we have $x_{i}=\partial_{i}\phi_{i}$,
		\begin{equation} 
			\begin{aligned}
				\int d\tau (\partial_{i}\phi_{i}(\tau)) \phi_{i'}(\tau)
				=i\delta_{ii'}+\int d\tau \phi_{i'}(\tau)(\partial_{i}\phi_{i}(\tau)),
			\end{aligned}
		\end{equation}
		with the delta-function in vacuum expectation-type definition:
		$i\delta_{ii'}=\frac{1}{i-i'-{\bf i}0^{+}}+\frac{1}{i'-i-i0^{+}}$.
In terms of density matrix and the correlation function in a basis of initial state $\psi$,
\begin{equation} 
	\begin{aligned}
		\frac{\delta \langle x_{j}\rangle_{\psi}}{\delta \phi_{i}}=
		\langle x_{i}x_{j}\rangle_{\psi}
		={\rm Tr}[
		x_{i}x_{i'}e^{\tau_{j}H_{j}}e^{-\tau_{j}H_{j}}]
		=\frac{{\rm Tr}[x_{i}e^{(\tau_{j}-\Delta \tau_{i})H_{j}}x_{i'}e^{-\tau_{j}H_{j}}]}
		{{\rm Tr}[e^{-\Delta \tau_{i}}]}
	\end{aligned}
\end{equation}
where we consider the a Hermitian case whose expectation is  evaluated through a microcanonical ensemble at thermal equilibrium $e^{\tau_{j}H_{j}}e^{-\tau_{j}H_{j}}$
which bring the corresponding microcanonical ensemble's character as carriered by $\tau_{j}$.
		
		The corresponding conservation can be  projected to the 
		exponential integral over the disordered Gaussian-distributed variables
		\begin{equation} 
			\begin{aligned}
				\label{7101}
				\int \mathcal{D}[\hat{\psi}] e^{-(\psi_{1}\psi_{2}-			\hat{\psi}_{1}\hat{\psi}_{2})}=1.
			\end{aligned}
		\end{equation}

		\begin{equation} 
			\begin{aligned}
				e^{-e^{a}e^{b}}=
				\int \mathcal{D}[\hat{\psi}] e^{-\hat{\psi}_{1}\hat{\psi}_{2}}
				e^{-\psi_{1}(\sum_{j}e^{b}_{j}x_{j})}
				e^{-\psi_{2}(\sum_{i}e^{a}_{i}x_{i})}.
			\end{aligned}
		\end{equation}
		Next we further inserting another set of identity
		in terms of the Gaussian distributed variables
		$(\sqrt{\hat{\psi}_{1}\hat{\psi}_{2}}+\overline{\psi}_{1})$ and
		$(\sqrt{\hat{\psi}_{1}\hat{\psi}_{2}}+\overline{\psi}_{2})$,
		\begin{equation} 
			\begin{aligned}
				2(\sqrt{\hat{\psi}_{1}\hat{\psi}_{2}}+\overline{\psi}_{1})\overline{\psi}_{1}+
				2(\sqrt{\hat{\psi}_{1}\hat{\psi}_{2}}+\overline{\psi}_{1})\overline{\psi}_{1}
				=\psi_{1}(\sum_{j}e^{b}_{j}x_{j})+\psi_{2}(\sum_{i}e^{a}_{i}x_{i}),
			\end{aligned}
		\end{equation}
		which indeed corresponds to the invariance provided by Eq.(\ref{7101}),
		\begin{equation} 
			\begin{aligned}
				\label{7102}
				\overline{(\sqrt{\hat{\psi}_{1}\hat{\psi}_{2}}+\overline{\psi}_{1})^2}
				=\overline{(\sqrt{\hat{\psi}_{1}\hat{\psi}_{2}}+\overline{\psi}_{2})^2}
				=\overline{(			\hat{\psi}_{1}\hat{\psi}_{2}-\psi_{1}\psi_{2})}.
			\end{aligned}
		\end{equation}
		Then we have
		\begin{equation} 
			\begin{aligned}
				e^{-e^{a}e^{b}}=
				\int \mathcal{D}			[\hat{\psi}] \int \mathcal{D}[\overline{\psi}]
				e^{-\hat{\psi}_{1}\hat{\psi}_{2}}
				e^{(\overline{\psi_{1}})^2}
				e^{(\overline{\psi_{2}})^2},
			\end{aligned}
		\end{equation}
		where using Eq.(\ref{7102}) we can obtain
		\begin{equation} 
			\begin{aligned}
				&
				\overline{-\psi_{1}\psi_{2}}
				=\overline{-(\overline{\psi}_{1})^2+\psi_{1}(\sum_{j}e^{b}_{j}x_{j})}
				=\overline{-(\overline{\psi}_{2})^2+\psi_{2}(\sum_{i}e^{a}_{i}x_{i})}.
			\end{aligned}
		\end{equation}
		
		In this way, the density matrix $h$ has been decouped, in terms of path integral representation, to actions including the  fermion Green function ($G$) and boson Green function ($D$):
		\begin{equation} 
			\begin{aligned}
				&
				G_{1}(\tau-\tau')=\overline{\psi}_{1}(\tau)\overline{\psi}_{1}(\tau'),\\
				&
				G_{2}(\tau-\tau')=\overline{\psi}_{2}(\tau)\overline{\psi}_{2}(\tau'),\\
				&
				\hat{D}(\tau-\tau')=\hat{\psi}_{1}(\tau)\hat{\psi}_{2}(\tau').
			\end{aligned}
		\end{equation}
Inserting the identity where the bosonic self-energy $\hat{\Sigma}$ plays the role of Lagrange multipliers
		\begin{equation} 
	\begin{aligned}
\int\mathcal{D}[\hat{\Sigma}]\mathcal{D}[\hat{D}]e^{-\int d\tau\int d\tau'\hat{\Sigma}(\tau-\tau')
	(\hat{D}(\tau-\tau')-\hat{\psi}_{1}(\tau)\hat{\psi}_{2}(\tau'))}=1,
	\end{aligned}
\end{equation}
whose static solutions are available through saddle-point approximation
		\begin{equation} 
			\begin{aligned}
				e^{-e^{a}e^{b}}&=
				\int \mathcal{D}			[\hat{\psi}] \int \mathcal{D}[\overline{\psi}]
				e^{-\int d\tau\int d\tau' \hat{\psi}_{1}(\tau)(\partial_{\tau}\delta_{\tau,\tau'}-\hat{\Sigma}
					)\hat{\psi}_{2}(\tau')}
				e^{-\hat{\Sigma} \hat{D}}
				e^{\overline{G}_{1}}
				e^{\overline{G}_{2}}
			\end{aligned}
		\end{equation}

	\renewcommand\refname{References}
	
	\clearpage

	\end{small}

\end{document}